\documentclass[a4paper, 11pt]{article}

\usepackage{xcolor}
\usepackage[margin=2.5cm]{geometry}

\definecolor{my_red}{RGB}{230,0,10}
\definecolor{my_blue}{RGB}{0,80,200}
\usepackage{hyperref}
\hypersetup{
    colorlinks    = true,
    linkcolor     = my_red,
    citecolor     = my_blue,
    urlcolor      = my_blue,
}
\usepackage[T1]{fontenc}
\usepackage[english]{babel}
\usepackage{csquotes}
\usepackage{tikz}
\usetikzlibrary{plotmarks}
\usetikzlibrary{external}
\usetikzlibrary{3d}
\usepackage[title]{appendix}
\usepackage{pgfplots}
\pgfplotsset{compat=1.17}
\usepackage[labelfont=sc, font=small]{caption}
\usepackage{subcaption}
\usepackage{float}
\newcommand{\deriv}{\mathrm{d}}
\newcommand{\vect}[1]{\boldsymbol{#1}}
\usepackage{dsfont}
\usepackage{algcompatible}
\usepackage{algorithm}
\usepackage{booktabs}
\usepackage{amsmath, amssymb, amsthm}
\newtheorem{remark}{Remark}[section]
\usepackage[shortlabels]{enumitem}
\usepackage{mathtools}
\mathtoolsset{showonlyrefs}
\usepackage[nottoc]{tocbibind}
\usepackage{xurl}
\usepackage{doi}

\usepackage[round]{natbib}
\title{Structure-preserving stochastic parameterization of a barotropic coupled ocean-atmosphere model with Ornstein--Uhlenbeck noise}
\author{Kamal Kishor Sharma\footnote{University of Hamburg, Germany (\href{mailto:kamal.sharma@uni-hamburg.de}{kamal.sharma@uni-hamburg.de})} \and
Peter Korn\footnote{Max Planck Institute for Meteorology, Germany}}

\begin{document}

\maketitle

\begin{abstract}
We present the first application of the stochastic advection by Lie transport (SALT) framework to an idealized coupled ocean-atmosphere system. SALT derives stochastic fluid equations from Hamilton's variational principle under a stochastic Lagrangian kinematic assumption, thereby preserving the geometric structure—Kelvin circulation theorem, Lie-derivative advection operators, and local conservation laws—of the underlying deterministic equations. The atmospheric component is rendered stochastic while the ocean remains deterministic, following Hasselmann's paradigm of a fast stochastic atmosphere driving a slow climate. The spatial correlation vectors encoding unresolved subgrid transport are estimated from high-resolution simulations via EOF analysis of Lagrangian trajectory differences. A central contribution is the replacement of the standard white-noise temporal model with Ornstein--Uhlenbeck (OU) processes, motivated by strong autocorrelation (decorrelation times of 50--150 time steps) in the dominant EOF modes; the OU process is the unique stationary Gaussian Markov process capable of capturing this temporal memory with a single parameter. Ensemble forecasts exhibit good spread--error agreement over 10--15 time units. Evaluated via the Continuous Ranked Probability Score—a strictly proper scoring rule measuring discrepancy between the forecast measure and the true conditional distribution—the stochastic ensemble consistently outperforms a size-matched deterministic ensemble, despite carrying higher RMSE. Well-posedness of the coupled stochastic system is identified as an important open problem.
\end{abstract}

\tableofcontents

%%%%%%%%%%%%%%%%%%%%%
\section{Introduction}
%%%%%%%%%%%%%%%%%%%%%

\subsection*{Geometric mechanics and stochastic fluid dynamics}

The equations governing geophysical fluid dynamics possess a rich geometric structure: they arise as Euler--Poincaré equations on the group of volume-preserving diffeomorphisms \citep{holmVariationalPrinciplesStochastic2015}, and this structure is the origin of their conservation laws—the Kelvin circulation theorem, potential vorticity conservation, and the Casimir invariants of the associated Lie--Poisson bracket. Any stochastic extension of these equations that discards this structure risks losing the dynamical constraints that make long-time statistical behavior physically meaningful.

Stochastic advection by Lie transport (SALT) \citep{holmVariationalPrinciplesStochastic2015} provides a route to stochastic fluid equations that preserves precisely this geometric content. Noise is introduced at the level of the Lagrangian fluid trajectory: the deterministic flow map $\mathbf{X}(t)$ is replaced by a stochastic flow $\mathbf{X}(t) + \sum_i \boldsymbol{\xi}_i \circ W^i_t$, where the $\boldsymbol{\xi}_i$ are divergence-free spatial vector fields encoding the statistics of unresolved transport and $W^i_t$ are independent Brownian motions. The stochastic model equations are then derived from Hamilton's variational principle under this modified kinematic assumption, so that the Kelvin circulation theorem, the Lie-derivative structure of the advection operators, and the local conservation laws are all inherited automatically from the derivation rather than imposed ad hoc. The resulting equations are SPDEs in Stratonovich form, a requirement for the chain rule of differential geometry to hold in the stochastic setting.

This structure has concrete analytical consequences. Well-posedness results for SALT equations in two spatial dimensions have been established for the incompressible Euler equations \citep{langWellposednessStochastic2D2023, flandoli2DEulerEquations2021} and for quasi-geostrophic models, relying critically on the geometric structure of the noise. The stochastic Kelvin theorem—stating that the circulation of the stochastic velocity around any material loop advected by the full stochastic flow is a martingale—holds for all SALT systems regardless of the specific fluid model, and provides a diagnostic for verifying that numerical discretizations respect the intended structure.

\subsection*{From geometry to ensemble forecasting}

The geometric structure of SALT has a direct practical payoff: because the noise is tied to the advection operator rather than added externally, the stochastic ensemble samples the uncertainty in the Lagrangian trajectories of the flow, which is precisely the uncertainty associated with unresolved subgrid transport. This makes SALT a natural framework for representing model error due to coarse graining.

Numerical implementations have been demonstrated for the 2D Euler equations \citep{cotterNumericallyModelingStochastic2019}, two-layer quasi-geostrophic models \citep{cotterModellingUncertaintyUsing2020}, shallow water equations \citep{crisanNoiseCalibrationSPDEs2023}, and surface quasi-geostrophic models \citep{resseguierDatadrivenSelfsimilarParameterizations2020}. Across these studies, SALT ensembles capture the true solution for a physically meaningful window while maintaining sufficient spread—a prerequisite for ensemble-based data assimilation \citep{cotterParticleFilterStochastic2020, cotterDataAssimilationQuasiGeostrophic2020, langBayesianInferenceFluid2022}.

A recurring simplification in the SALT calibration pipeline is that the temporal coefficients of the noise modes are modeled as independent Gaussian draws, implicitly assuming that unresolved transport has no temporal memory. \citet{ephratiDataDrivenStochasticLie2023} showed, for the stochastic Euler equations, that this assumption is violated in practice: the leading EOF modes exhibit strong temporal autocorrelation, and replacing Gaussian noise with Ornstein--Uhlenbeck (OU) processes—which possess finite decorrelation time and arise naturally as the stationary limit of a colored-noise approximation to Brownian motion—markedly improves ensemble reliability. The OU process is also the unique Gaussian Markov process that is stationary and mean-reverting, making it the minimal extension of white noise that can capture temporal memory without introducing additional parameters beyond the decorrelation time.

\subsection*{Contributions of this paper}

The present paper makes two primary contributions. First, we carry out the first application of SALT to an idealized coupled ocean-atmosphere climate model, extending the framework from single-component geophysical systems to a genuinely coupled setting in which the fast stochastic atmosphere drives a slow deterministic ocean consistent with Hasselmann's paradigm \citep{hasselmannStochasticClimateModels1976}. Second, we show that OU-process noise—calibrated via an AR(1) model fitted to the measured autocorrelation of each EOF mode—consistently outperforms white-noise calibration, and that the resulting stochastic ensemble dominates a size-matched deterministic ensemble under the Continuous Ranked Probability Score (CRPS), a strictly proper scoring rule that simultaneously rewards both accuracy and calibration of the forecast distribution.

\subsection{The climate model}\label{subsec: climate model intro}

We numerically solve the idealized coupled ocean-atmosphere model of \citet{crisanImplementationHasselmannsParadigm2023}. Following Hasselmann's paradigm \citep{hasselmannStochasticClimateModels1976, lucariniTheoreticalToolsUnderstanding2023}, the fast atmospheric component is rendered stochastic while the slow ocean remains deterministic. The deterministic equations read
\begin{align}\label{eq: non dim det climate}
    \text{Atmosphere}: \quad &\frac{\partial \mathbf{u}^a}{\partial t} + (\mathbf{u}^a\cdot \nabla)\mathbf{u}^a + \frac{1}{Ro^a} \hat{\mathbf{z}} \times \mathbf{u}^a + \frac{1}{C^a} \nabla \theta^a = {\nu}^a \Delta \mathbf{u}^a, \\
    &\frac{\partial \theta^a}{\partial t} + \nabla \cdot (\mathbf{u}^a \theta^a) = \gamma(\theta^a - \theta^o) + \eta^a \Delta \theta^a,\\
    \text{Ocean}: \quad &\frac{\partial \mathbf{u}^o}{\partial t} + (\mathbf{u}^o\cdot \nabla)\mathbf{u}^o + \frac{1}{Ro^o} \hat{\mathbf{z}} \times \mathbf{u}^o + \frac{1}{Ro^o} \nabla p^o = \sigma(\mathbf{u}^o - \mathbf{u}_{sol}^a) + \nu^o \Delta \mathbf{u}^o,\\
    & \nabla \cdot \mathbf{u}^o = 0, \qquad \frac{\partial \theta^o}{\partial t} + \mathbf{u}^o \cdot \nabla \theta^o = \eta^o \Delta \theta^o.
\end{align}
Here $\mathbf{u}^{a/o}$, $\theta^{a/o}$, $p^o$ are the velocity, temperature, and pressure fields for the atmosphere and ocean, respectively, on the 2D domain $\vect{x}=(x,y)$. The atmosphere is governed by 2D compressible Navier--Stokes equations coupled to an advection--diffusion equation for temperature; thermal coupling enters through $\gamma(\theta^a - \theta^o)$ and wind-stress coupling through $\sigma(\mathbf{u}^o - \mathbf{u}^a_{sol})$, where $\mathbf{u}^a_{sol}$ is the divergence-free part of $\mathbf{u}^a$. The non-dimensional parameters are
$$Ro^a = Ro^o = \frac{U}{Lf}, \quad C^a = \frac{U^2}{\kappa \Theta}, \quad \kappa = c_v \!\left(\frac{R}{p_0}\right)^{\!2/7},$$
with $\Theta$ a reference temperature, $f$ the Coriolis parameter, and $\nu^{a/o}$, $\eta^{a/o}$ resolution-dependent eddy viscosity and diffusivity coefficients.

Applying the SALT derivation to the atmospheric component yields the stochastic climate model
\begin{align}\label{eq: non dim stoch climate}
\text{Atmosphere}: \quad &\deriv \mathbf{u}^a + \!\left(\mathbf{u}^a \deriv t + \sum_i \boldsymbol{\xi}_i \circ \deriv W^i_t\right)\!\cdot \nabla \mathbf{u}^a + \frac{1}{Ro^a} \hat{\mathbf{z}} \times \mathbf{u}^a\deriv t \nonumber \\
                &\quad + \sum_i (u_1^a \nabla \xi_{i,1} + u_2^a \nabla \xi_{i,2} )\circ \deriv W^i_t = \!\left(-\frac{1}{C^a} \nabla \theta^a + \nu^a \Delta \mathbf{u}^a\right)\! \deriv t, \\
        &\deriv \theta^a + \nabla\cdot (\theta^a \mathbf{u}^a)\deriv t + \sum_i (\boldsymbol{\xi}_i \circ \deriv W^i_t) \cdot \nabla \theta^a = (\gamma(\theta^a - \theta^o) + \eta^a \Delta \theta^a )\deriv t,\\
\text{Ocean}: \quad & \text{(unchanged, except } \mathbf{u}^a_{sol} \text{ replaced by } \mathbb{E}\mathbf{u}^a_{sol} \text{ in coupling term)},
\end{align}
where $W_t^i$ are independent Brownian motions and $\boldsymbol{\xi}_i$ are spatial correlation vectors encoding the statistics of unresolved transport (see Appendix~\ref{app: cali}). The Stratonovich form $\circ$ preserves the Lie-algebraic structure of the original equations. Derivation details and the non-dimensionalization are given in Appendix~\ref{app: nondim}.

\begin{remark}[Well-posedness]
\normalfont
Well-posedness of the SALT system \eqref{eq: non dim stoch climate} in the fully coupled setting is, to our knowledge, an open problem. For the incompressible SALT Euler equations in two dimensions, existence and uniqueness results have been established \citep{langWellposednessStochastic2D2023, flandoli2DEulerEquations2021}, relying on the geometric structure of the noise to obtain energy estimates. The compressibility of the atmospheric component and the two-way thermal coupling introduce additional technical difficulties that fall outside the scope of these results. For the deterministic system \eqref{eq: non dim det climate}, local well-posedness follows from standard energy methods for coupled Navier--Stokes systems. A rigorous treatment of the stochastic coupled problem—existence, uniqueness, and continuous dependence on initial data for \eqref{eq: non dim stoch climate}—remains an important open direction at the intersection of stochastic analysis and geophysical fluid dynamics. In the present work we proceed numerically and rely on ensemble consistency (Section~\ref{subsec: UQ}) as a practical diagnostic in lieu of analytical well-posedness.
\end{remark}

%%%%%%%%%%%%%%%%%%%%%
\section{Numerical simulations}\label{sec: numerical sim}
%%%%%%%%%%%%%%%%%%%%%

Simulations are performed on a rectangular domain $\Omega = [0,7]\times[0,1]$ (periodic in $x$), approximating atmospheric and oceanic dynamics between $27.5^\circ$ and $62.5^\circ$ N (physical dimensions $27{,}237\,\text{km} \times 3{,}891\,\text{km}$). Geophysical scaling yields $Ro^a = Ro^o = 0.3$ and $C^a = 0.02$. Free-slip and insulated Neumann conditions are imposed on the meridional boundaries. A finite element discretization is employed throughout (see Appendix~\ref{app: discretization}). Grid parameters are listed in Table~\ref{tab: parameter clima}.

\begin{table}[h]
    \centering
    \caption{Grid parameters for the atmosphere and ocean components.}
    \resizebox{.72\textwidth}{!}{%
    \begin{tabular}{l c c}
    \toprule
        \textbf{Parameter} & \textbf{Fine grid} & \textbf{Coarse grid}\\
        \midrule
         Elements $N_x \times N_y$ & $896 \times 128$ & $224 \times 32$ \\
         Smallest element $\Delta x$ & $1/128\ (\sim 30\ \text{km})$ & $1/32\ (\sim 120\ \text{km})$ \\
         Time step $\Delta t$ & $0.010\ (\sim 8\ \text{min})$ & $0.04\ (\sim 32\ \text{min})$ \\
         Eddy viscosity $\nu^{a/o}$, diffusivity $\eta^{a/o}$ & $1/(8\times10^4)$ & $1/10^4$ \\
         \bottomrule
    \end{tabular}}
    \label{tab: parameter clima}
\end{table}

\subsection{High-resolution reference simulation}\label{subsec: det model sim}

The deterministic model \eqref{eq: non dim det climate} is integrated on the fine grid from $t=0$ to $t=45$. The atmosphere is initialized with a smooth zonal jet
\begin{equation}
    u_1^a(x,y,0) = \begin{cases}
       0 & y \leq y_0 \text{ or } y \geq y_1, \\
       \exp\!\left(\dfrac{\alpha^2}{(y-y_0)(y-y_1)}\right)\exp\!\left(\dfrac{4\alpha^2}{(y_1-y_0)^2}\right) & y_0 < y < y_1,
     \end{cases}
\end{equation}
with $\alpha = 1.64$, $y_0 = 1/14$, $y_1 = 13/14$, and temperature in thermal-wind balance. The ocean starts at rest with a localized temperature perturbation. Coupling coefficients are $\gamma = -10$ and $\sigma = -0.1$.

\begin{figure}[h]
    \centering
    \includegraphics[width=.67\linewidth]{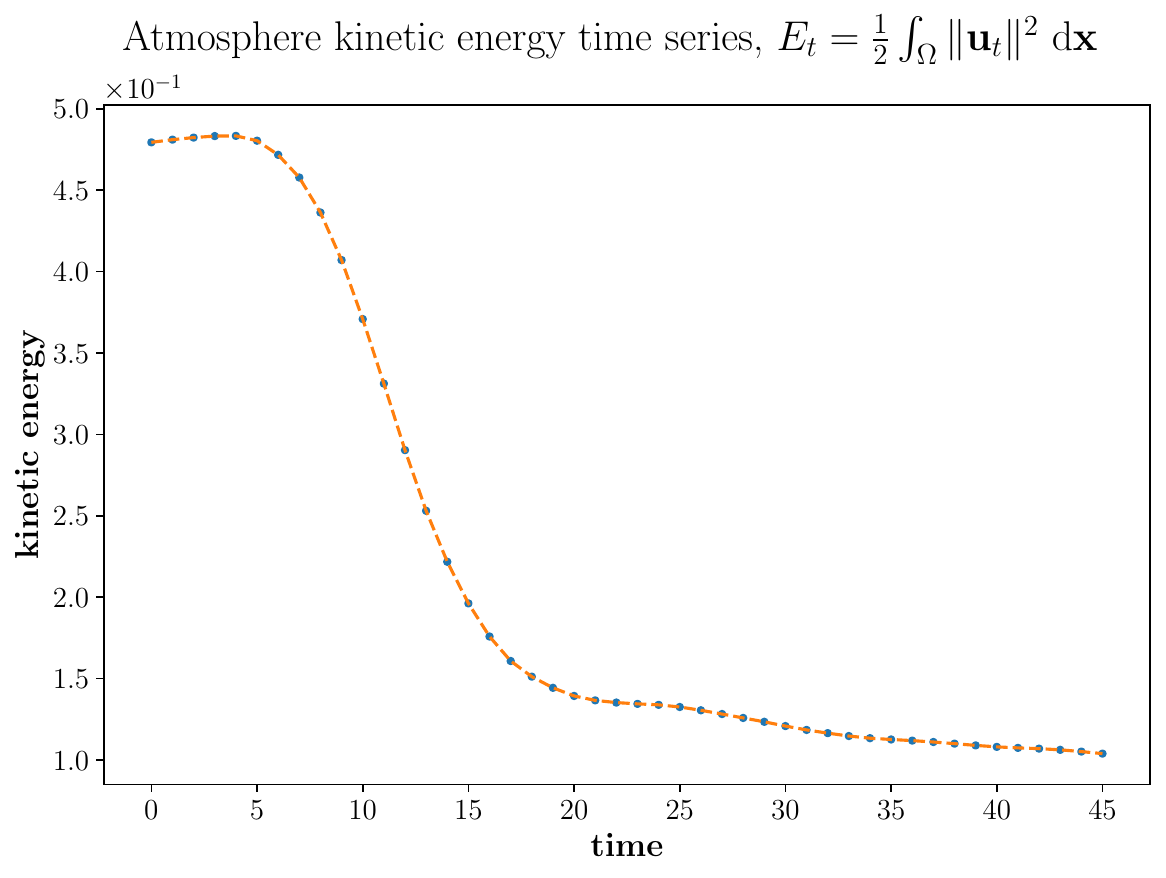}
    \caption{Time series of atmospheric kinetic energy. Statistical equilibrium is reached around $t=25$.}
    \label{fig: KE coup det}
\end{figure}

\begin{figure}
\centering
    \begin{subfigure}{1\textwidth}\centering
      \includegraphics[width=1\textwidth]{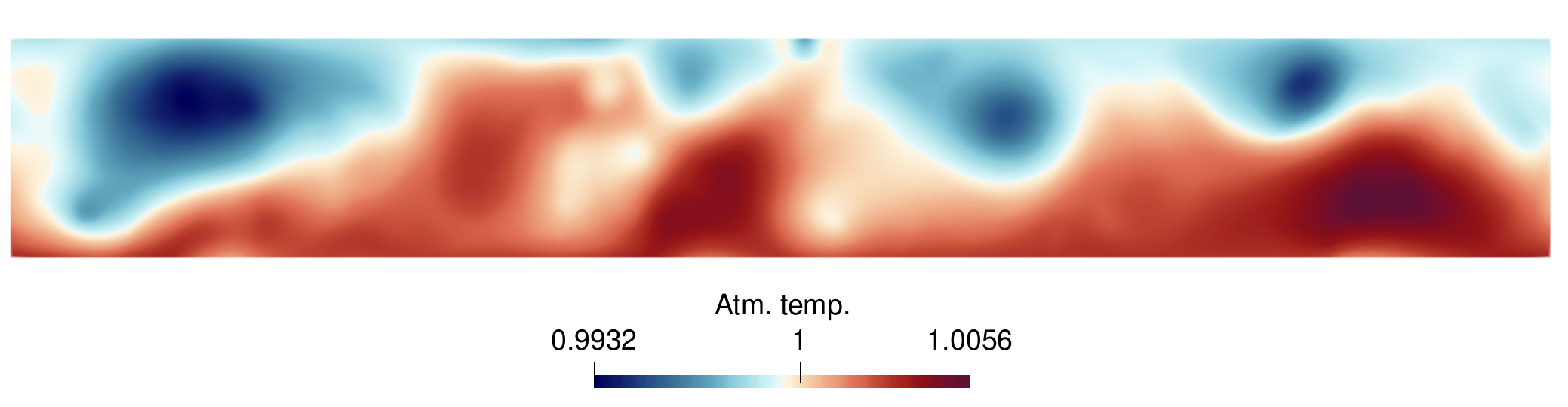}
    \end{subfigure}\par\medskip
    \begin{subfigure}{1\textwidth}\centering
      \includegraphics[width=1\textwidth]{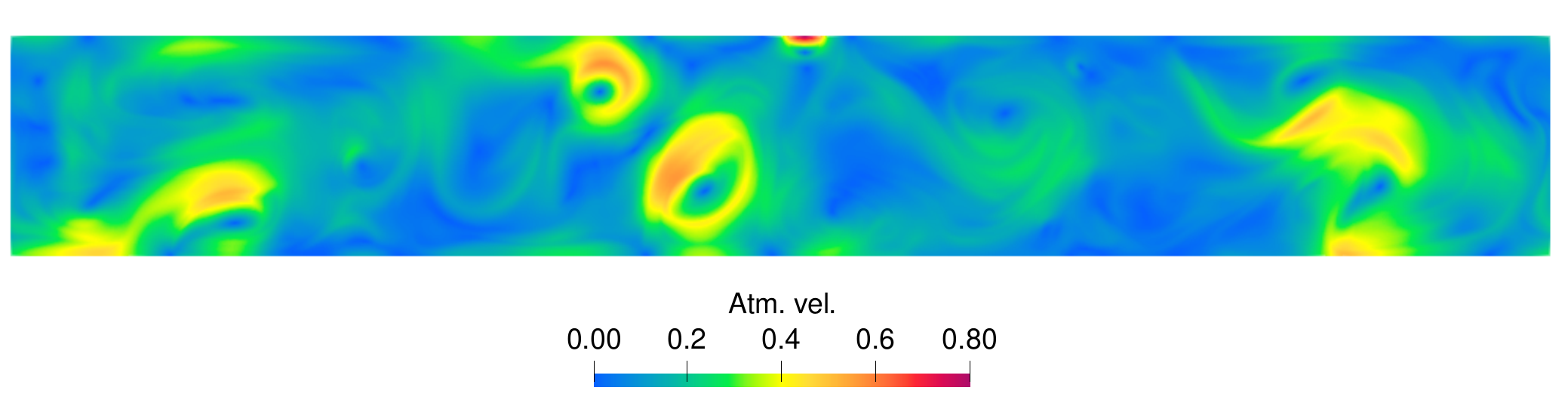}
    \end{subfigure}\par\medskip
    \begin{subfigure}{1\textwidth}\centering
      \includegraphics[width=1\textwidth]{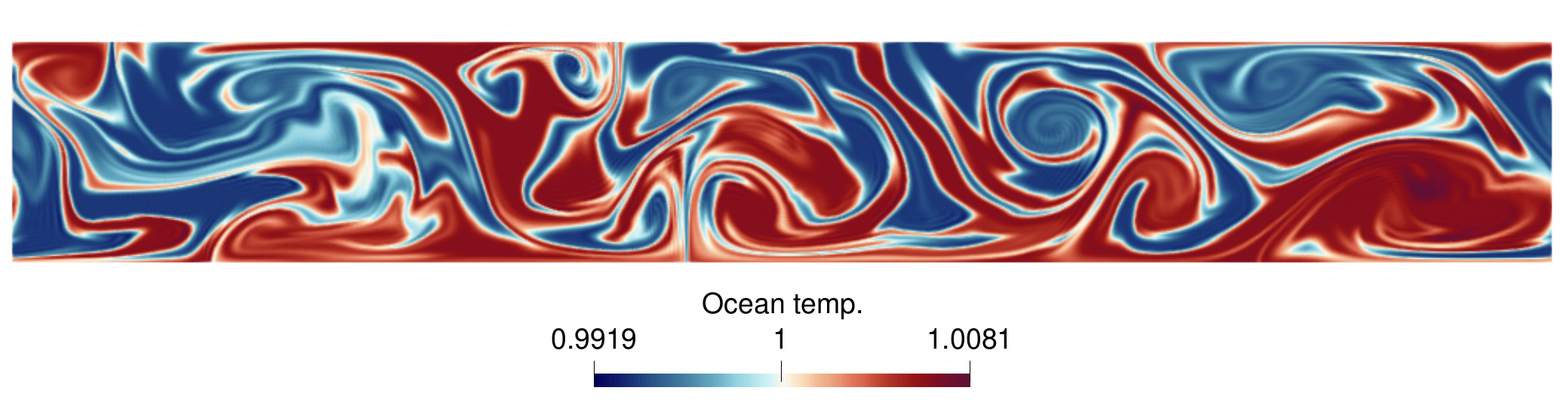}
    \end{subfigure}\par\medskip
    \begin{subfigure}{1\textwidth}\centering
      \includegraphics[width=1\textwidth]{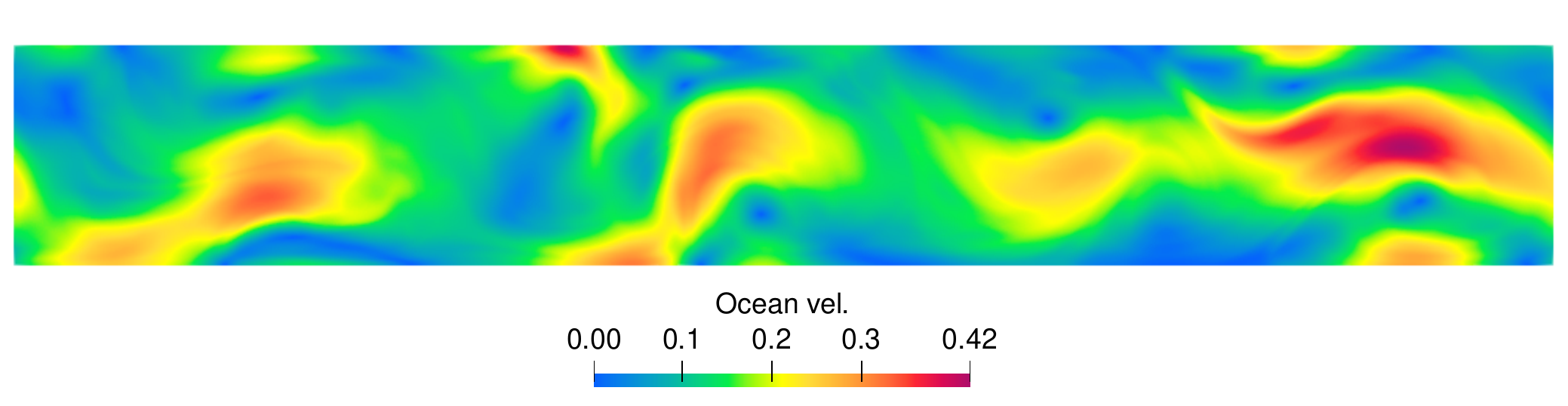}
    \end{subfigure}
    \caption{Atmosphere (top two panels) and ocean (bottom two panels) fields at $t=25$, showing multi-scale eddy structure and the onset of statistical equilibrium.}
    \label{fig: coup model t25}
\end{figure}

\begin{figure}
    \captionsetup[subfigure]{labelformat=empty}
\centering
    \begin{subfigure}{1\textwidth}\centering\caption{$t=25$}
      \vspace{-.1cm}
      \includegraphics[width=1\textwidth]{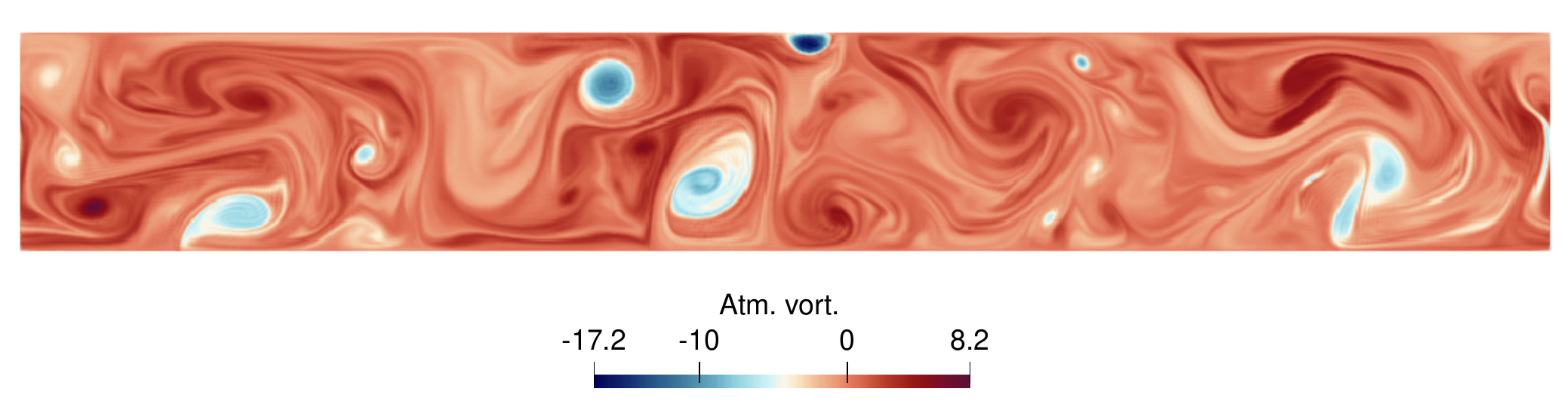}
    \end{subfigure}\par
    \begin{subfigure}{1\textwidth}\centering\caption{$t=30$}
      \vspace{-.1cm}
      \includegraphics[width=1\textwidth]{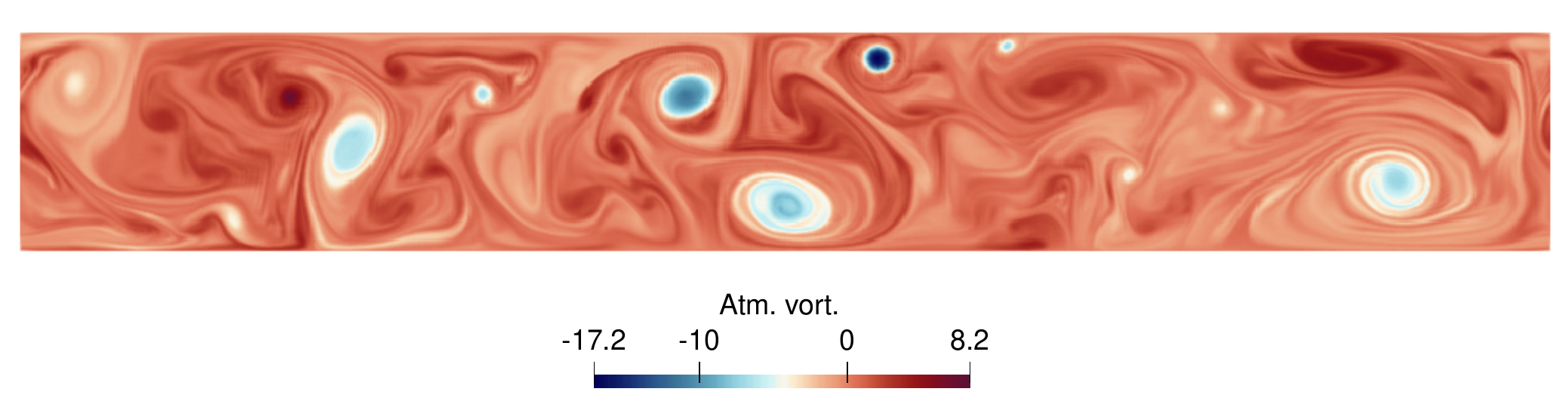}
    \end{subfigure}\par
    \begin{subfigure}{1\textwidth}\centering\caption{$t=35$}
      \vspace{-.1cm}
      \includegraphics[width=1\textwidth]{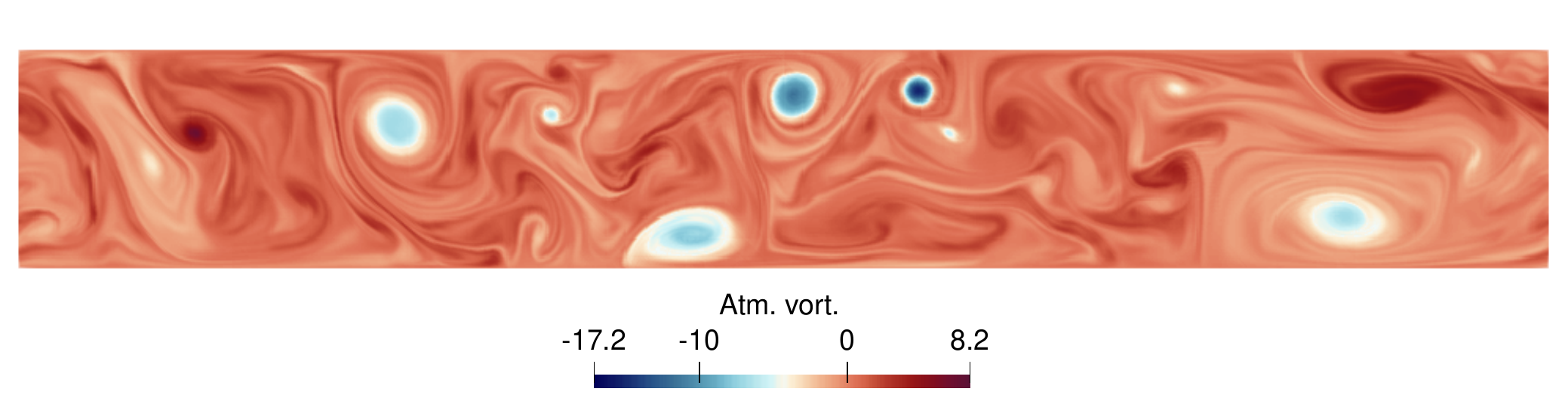}
    \end{subfigure}\par
    \begin{subfigure}{1\textwidth}\centering\caption{$t=40$}
      \vspace{-.1cm}
      \includegraphics[width=1\textwidth]{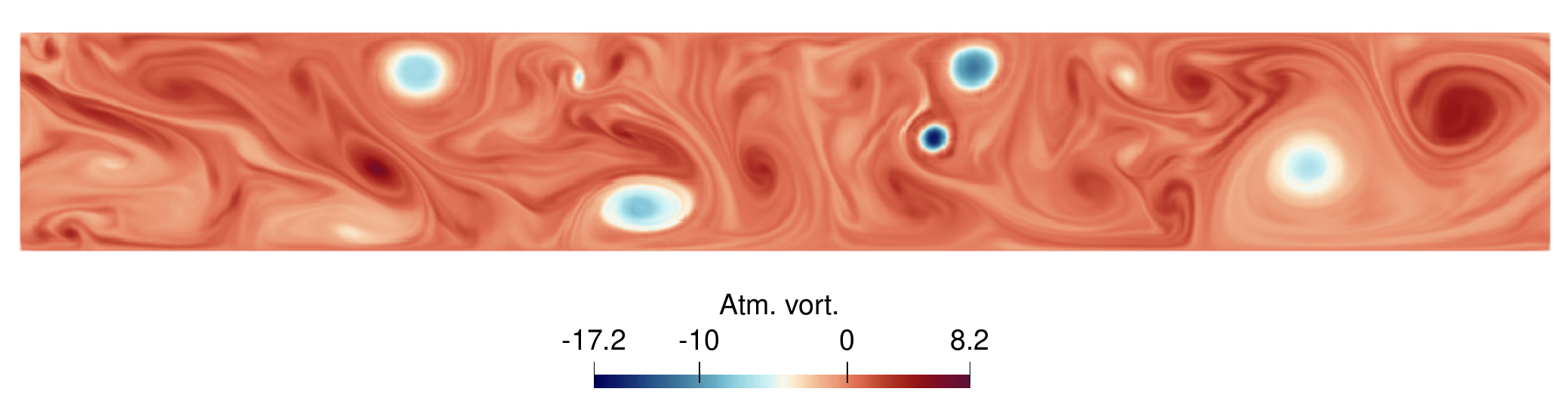}
    \end{subfigure}\par
    \begin{subfigure}{1\textwidth}\centering\caption{$t=45$}
      \vspace{-.1cm}
      \includegraphics[width=1\textwidth]{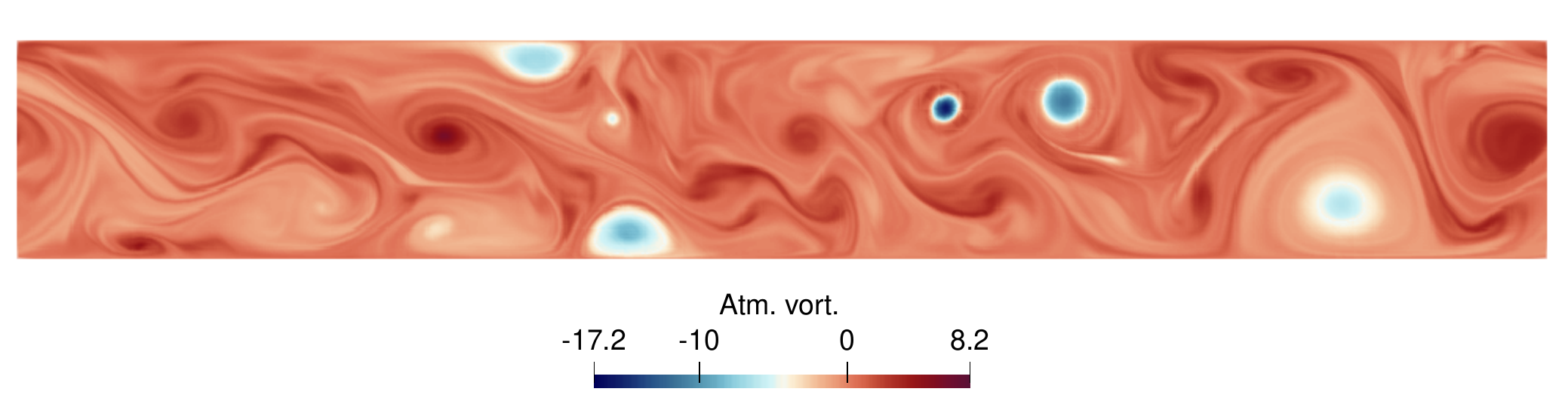}
    \end{subfigure}
    \caption{Atmospheric vorticity during the equilibrium window ($t=25$--$45$). Eddy counts and structure remain quasi-stationary, confirming suitability for calibration.}
    \label{fig: cm atm vort evo}
\end{figure}

After a spin-up of approximately 20 time units, the atmospheric kinetic energy stabilizes (Figure~\ref{fig: KE coup det}), signaling statistical equilibrium. Snapshots at $t = 25$--$45$ (Figures~\ref{fig: coup model t25}--\ref{fig: cm atm vort evo}) show a dynamically rich multi-scale flow that is quasi-stationary in a statistical sense. The velocity data from this equilibrium window are used for stochastic model calibration.

\subsection{Model calibration and OU noise}\label{subsec: model calib}

The spatial correlation vectors $\boldsymbol{\xi}_i$ are estimated from the Lagrangian trajectory difference $\mathbf{u} - \overline{\mathbf{u}}$, where $\overline{\mathbf{u}}$ is the Helmholtz-filtered, coarse-grained high-resolution velocity (see Appendix~\ref{app: cali}). SVD of the data matrix $(\mathbf{u}-\overline{\mathbf{u}})\sqrt{\Delta t}$ yields
\begin{equation}\label{eq: EOF algo}
    \mathbf{g}(\vect{x},t) = \boldsymbol{\xi}_0 + \sum_{i=1}^N {a}_i(t)\,\boldsymbol{\xi}_i(\vect{x}),
\end{equation}
where $\boldsymbol{\xi}_0$ is the time mean, $\boldsymbol{\xi}_i$ are the spatial empirical orthogonal function (EOF) modes, and $a_i(t)$ are the corresponding time series. The first 53, 20, and 10 EOFs explain 99\%, 90\%, and 70\% of total variance, respectively; we retain 53 EOFs for all stochastic simulations.

\begin{figure}
    \centering
    \includegraphics[width=1\linewidth]{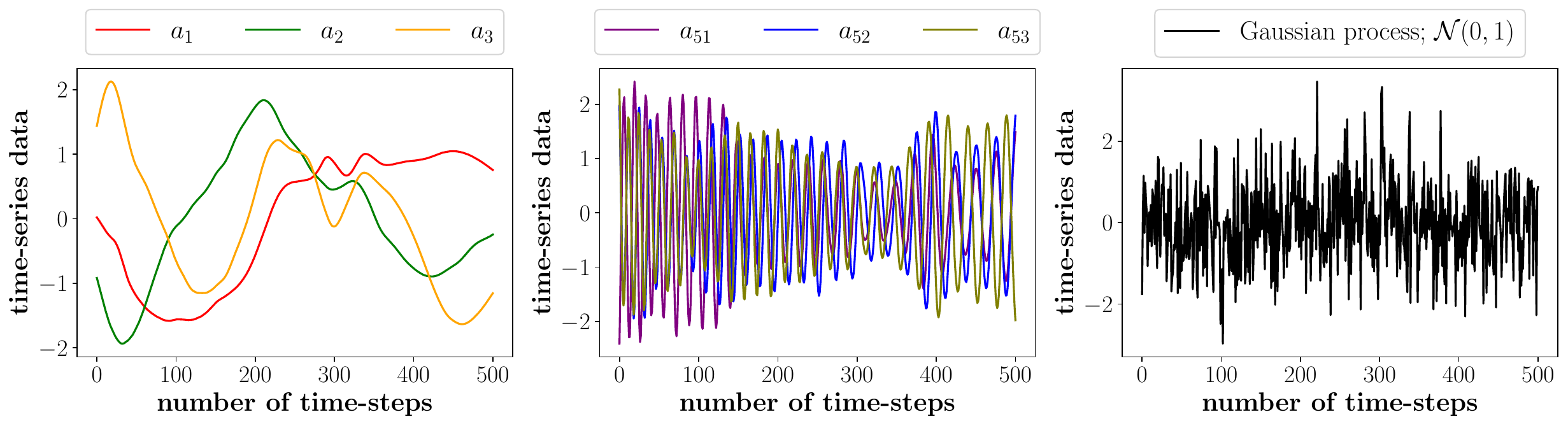}
    \caption{Time series $a_i(t)$ for the leading modes ($\boldsymbol{\xi}_{1}$--$\boldsymbol{\xi}_{3}$, left) and high modes ($\boldsymbol{\xi}_{51}$--$\boldsymbol{\xi}_{53}$, centre), compared with a Gaussian process realization (right). Leading modes are strongly autocorrelated.}
    \label{fig: time series a mesh 32 CM}
\end{figure}

\begin{figure}
    \centering
    \includegraphics[width=1\linewidth]{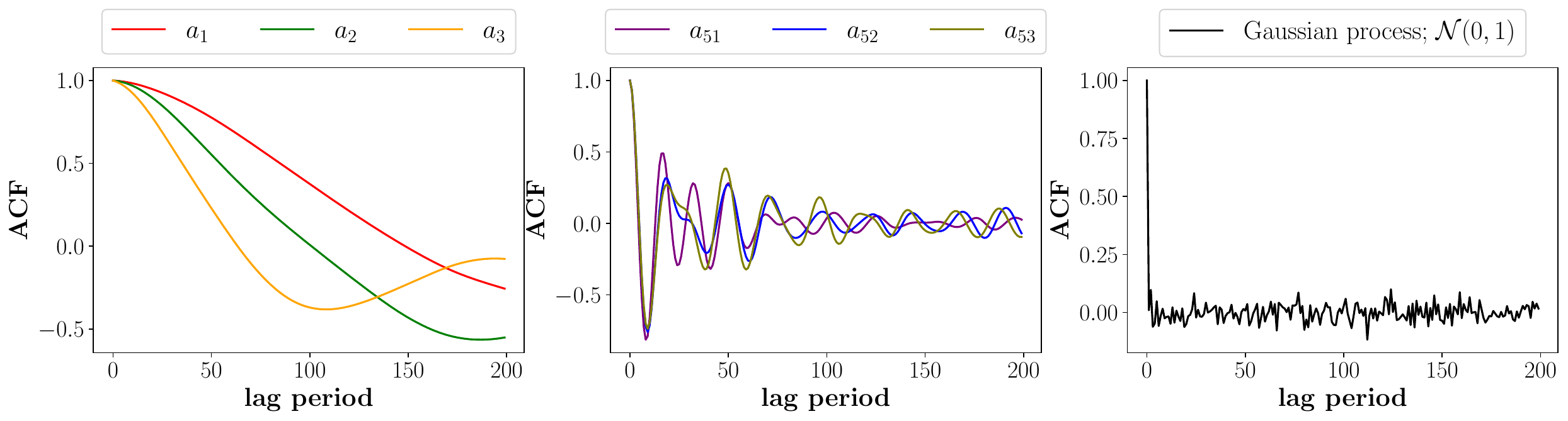}
    \caption{Autocorrelation functions of $a_i(t)$: leading modes (left) show decorrelation times of 50--150 steps; high modes (centre) and Gaussian noise (right) decorrelate within a single step.}
    \label{fig: acf xi mesh 32 CM}
\end{figure}

\paragraph{OU versus Gaussian noise.}
Standard practice approximates $a_i(t)$ as i.i.d.\ $\mathcal{N}(0,1)$ draws. However, autocorrelation function (ACF) analysis reveals that the dominant modes are strongly autocorrelated: the decorrelation time of $a_1$--$a_3$ is 50--150 time steps (Figures~\ref{fig: time series a mesh 32 CM}--\ref{fig: acf xi mesh 32 CM}), whereas white noise has decorrelation time 1. Imposing a Gaussian approximation discards this structure.


We therefore model each $a_i(t)$ with an Ornstein--Uhlenbeck process, discretized as an AR(1) process $X_{t} = \varphi X_{t-1} + \varepsilon_t$, where $\varphi = \operatorname{Cov}(a_i(t_k), a_i(t_{k+1}))/\operatorname{Var}(a_i)$ and $\varepsilon_t \sim \mathcal{N}(0, 1-\varphi^2)$ (Algorithm~\ref{alg: AR1}, Appendix~\ref{app: AR1 model}). Comparison experiments confirm that OU-calibrated ensembles outperform Gaussian-noise ensembles and are free of spurious oscillations \citep{sharma2025development}.

\subsection{Stochastic model simulation}\label{subsec: stoch model sim}

\begin{figure}
\centering
    \begin{subfigure}{1\textwidth}\centering
      \includegraphics[width=1\textwidth]{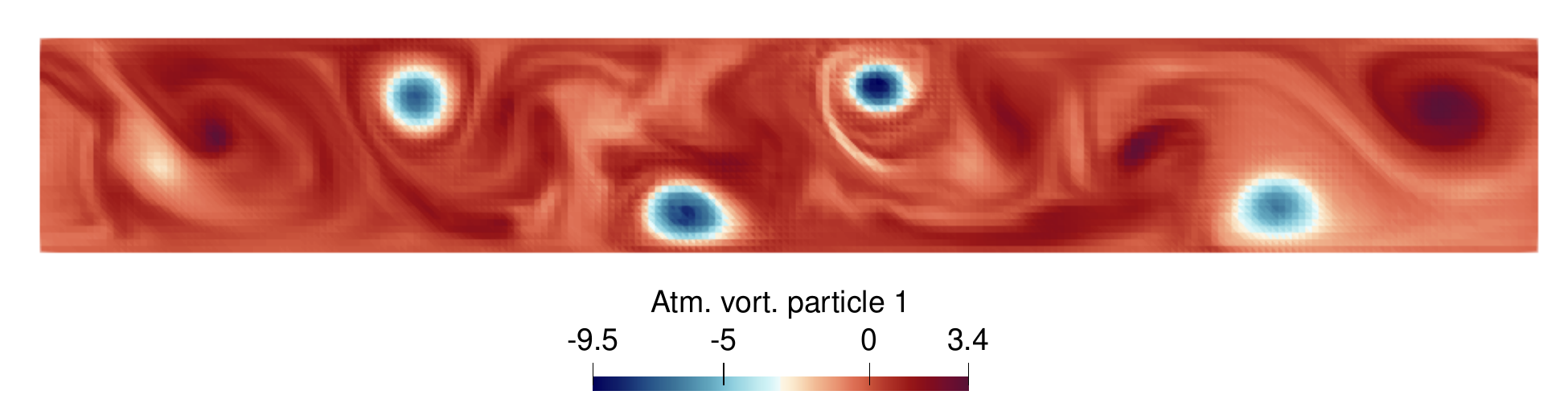}
    \end{subfigure}
    \begin{subfigure}{1\textwidth}\centering
      \includegraphics[width=1\textwidth]{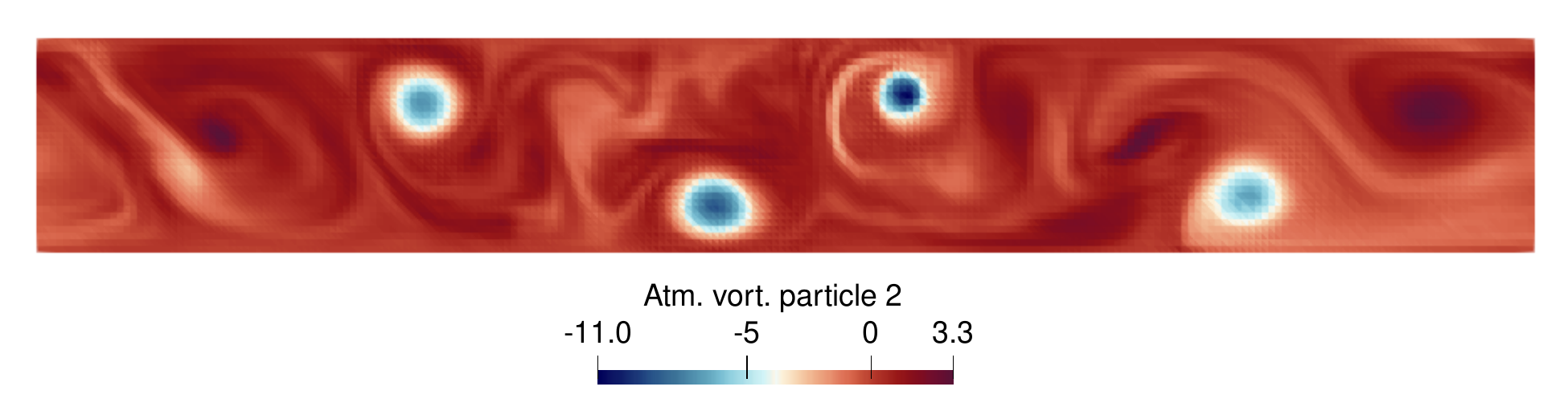}
    \end{subfigure}
    \begin{subfigure}{1\textwidth}\centering
      \includegraphics[width=1\textwidth]{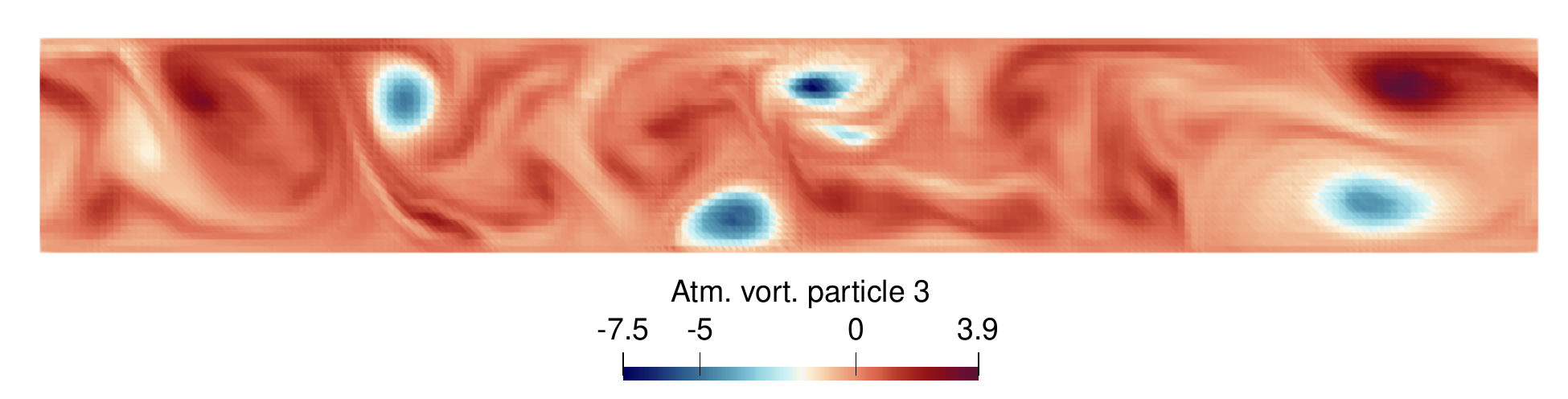}
    \end{subfigure}
    \begin{subfigure}{1\textwidth}\centering
      \includegraphics[width=1\textwidth]{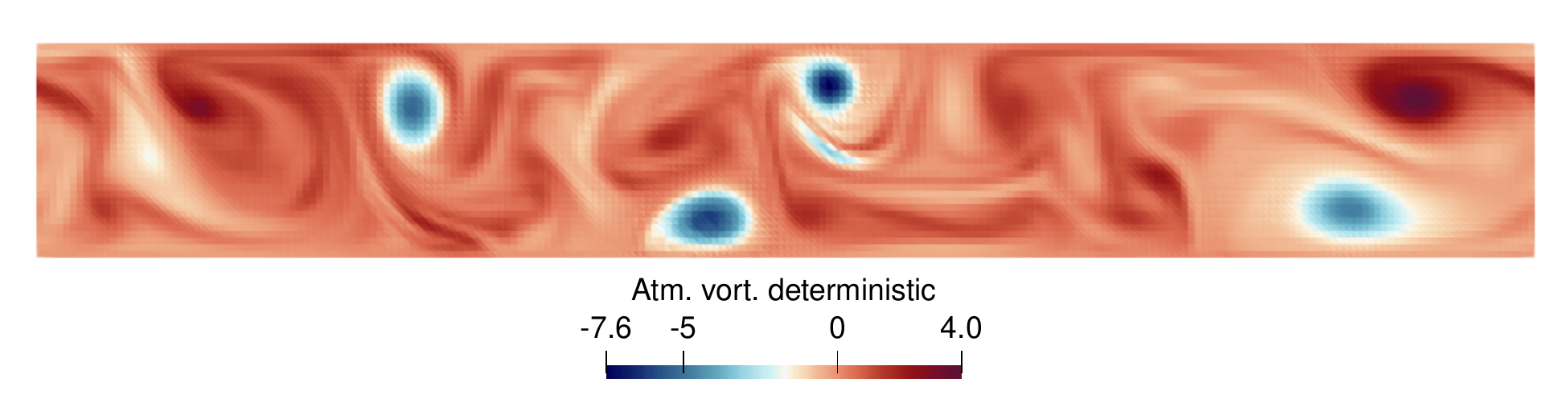}
    \end{subfigure}
    \begin{subfigure}{1\textwidth}\centering
      \includegraphics[width=1\textwidth]{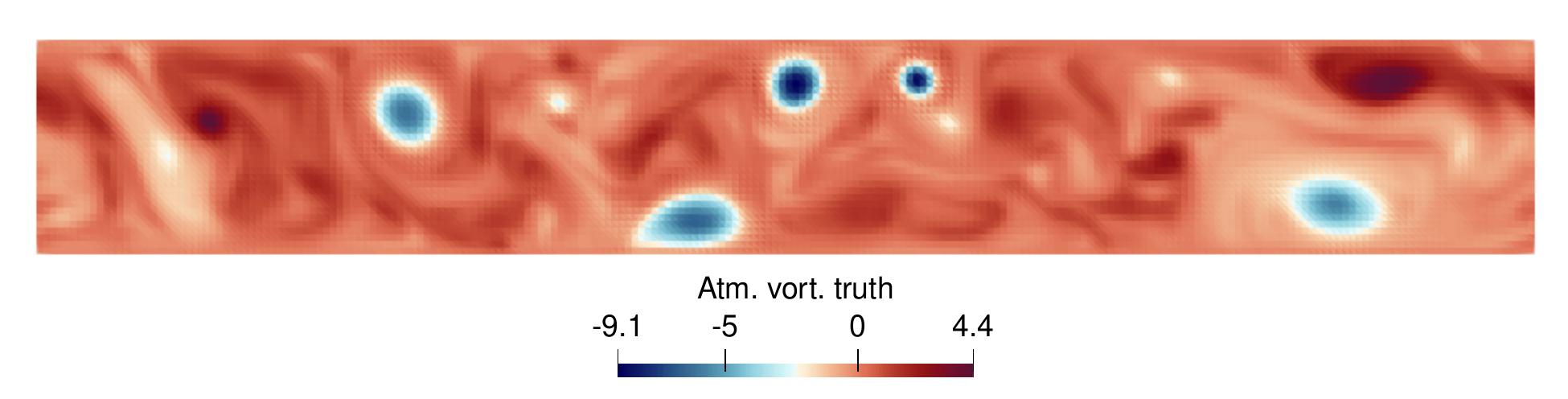}
    \end{subfigure}
    \caption{Atmospheric vorticity at $t=35$: three SPDE realizations (top three), deterministic coarse-grid solution (fourth), and truth (bottom). Stochastic particles exhibit higher spread in the dynamically uncertain central region.}
    \label{fig: ensem vort v det v truth t35}
\end{figure}

\begin{figure}
    \centering
    \includegraphics[width=1\linewidth]{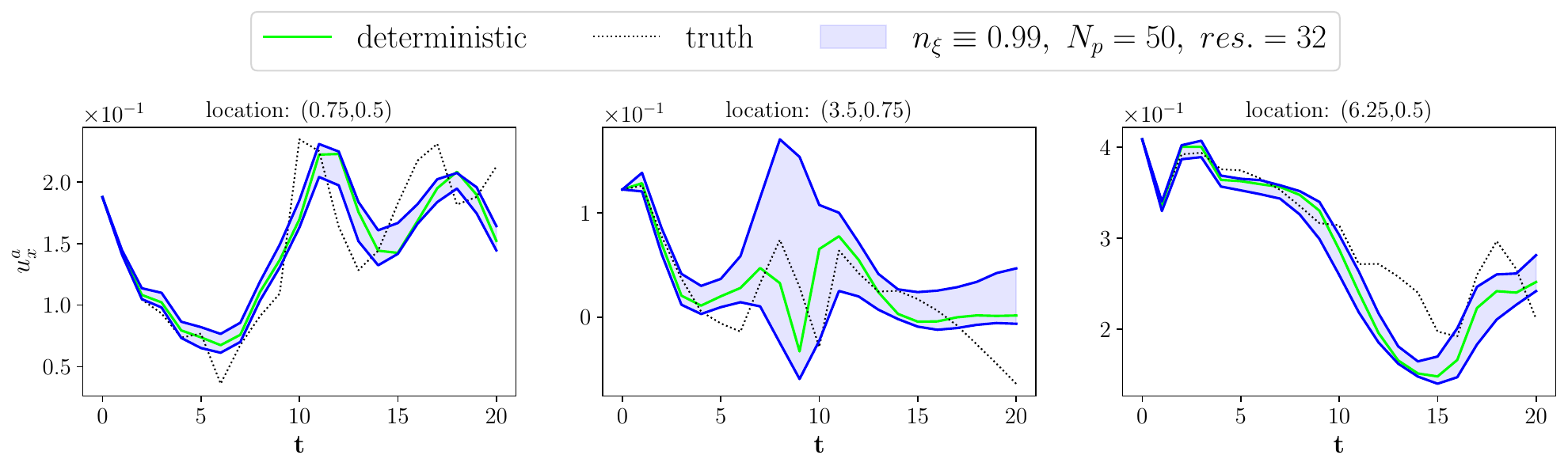}
    \caption{Evolution of atmospheric velocity ($x$ component) at three locations on the grid over time ($t=25$--$45$). One-standard-deviation band above the ensemble mean of stochastic model solution is compared with the coarse-grained high-resolution solution (truth) and the deterministic model solution.}
    \label{fig: three loc uq det v sto}
\end{figure}

The stochastic model \eqref{eq: non dim stoch climate} is run on the coarse grid ($224 \times 32$), initialized from coarse-grained fields at $t=25$, and integrated for 20 time units. An ensemble of $N_p = 50$ independent particles, all starting from the same initial condition, is evolved in parallel. Figure~\ref{fig: ensem vort v det v truth t35} shows ensemble vorticity realizations at $t=35$ alongside the deterministic coarse-grid solution and the coarse-grained high-resolution truth. The stochastic particles exhibit correct large-scale structure everywhere except the central region, where natural flow uncertainty is reflected by elevated inter-particle spread—a qualitatively informative signal absent from the single deterministic trajectory. This is further confirmed from the velocity plots (Figure~\ref{fig: three loc uq det v sto}) which compare the stochastic and deterministic solutions with the true solution at three locations: left plot corresponds to the coordinate $(0.75, 0.5)$, the center plot corresponds to $(3.5, 0.75)$, and the right plot corresponds to $(6.25, 0.5)$. Velocity fields exhibits considerably higher spread in the central region in comparison to other regions. 

\begin{figure}
\centering
    \begin{subfigure}{1\textwidth}\centering
      \includegraphics[width=1\textwidth]{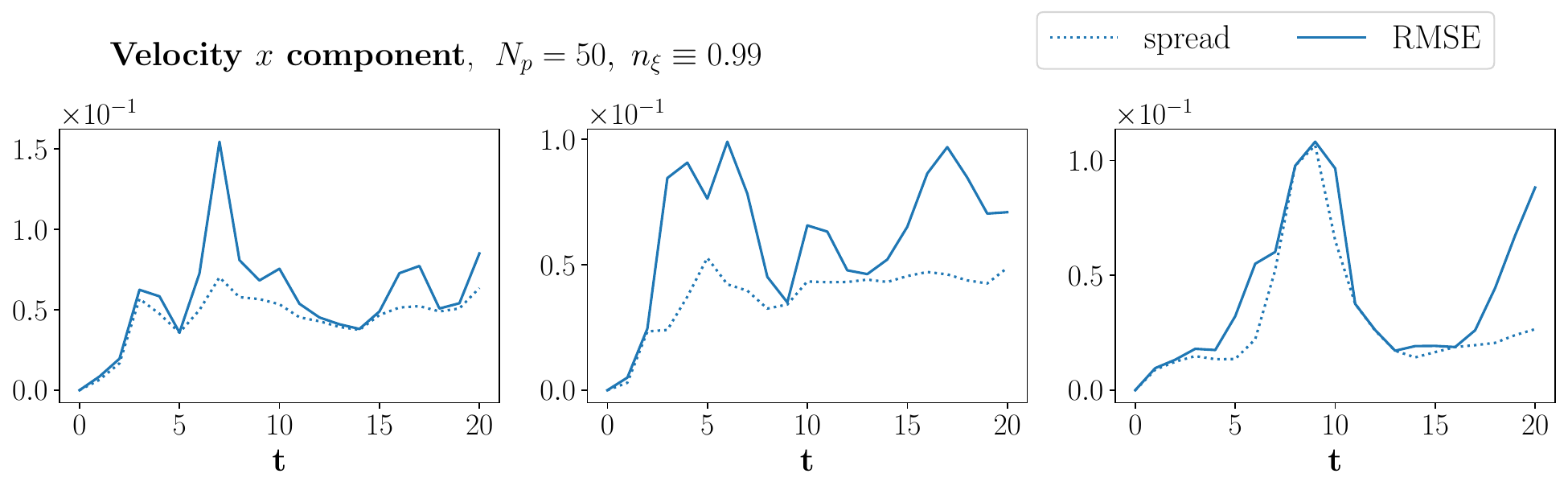}
    \end{subfigure}\par\medskip
    \begin{subfigure}{1\textwidth}\centering
      \includegraphics[width=1\textwidth]{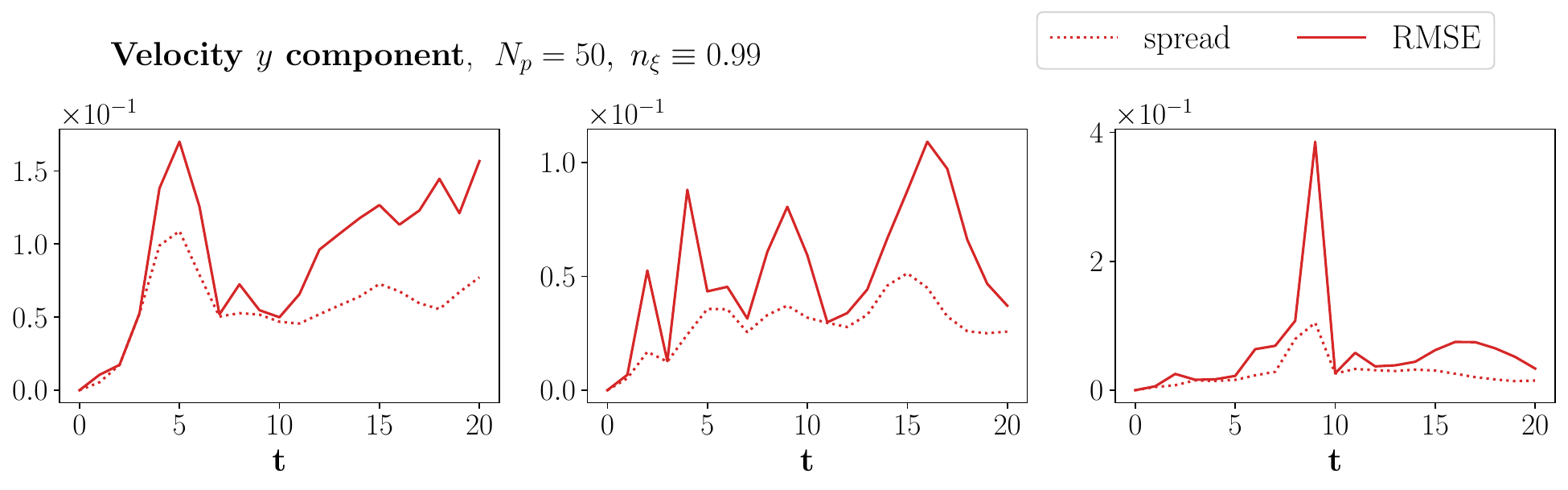}
    \end{subfigure}\par\medskip
    \begin{subfigure}{1\textwidth}\centering
      \includegraphics[width=1\textwidth]{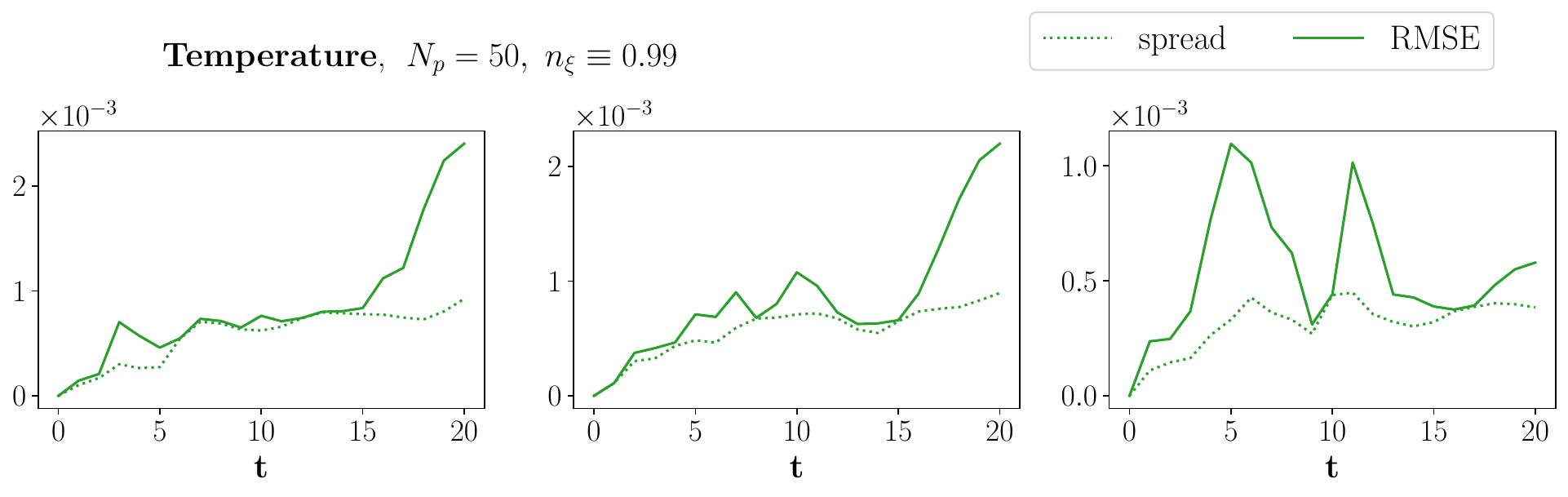}
    \end{subfigure}
    \caption{Ensemble spread and RMSE for atmospheric velocity ($u^a_x$, $u^a_y$) and temperature at three observation points ($N_p = 50$, $n_\xi \equiv 0.99$). Spread tracks RMSE for $\sim$15 time units.}
    \label{fig: vel temp rmse spread stoc}
\end{figure}

To quantify ensemble reliability, velocity and temperature are monitored at 84 uniformly distributed observation points. The ensemble mean, spread $\operatorname{Spread}(X_i) = [N_p^{-1}\sum_i(X_i - \hat{\mathbb{E}}X_i)^2]^{1/2}$, and root mean square error (RMSE) $\operatorname{RMSE}(X_i, X_\text{truth}) = [N_p^{-1}\sum_i(X_i - X_\text{truth})^2]^{1/2}$ are tracked over time. Figure~\ref{fig: vel temp rmse spread stoc} shows spread and RMSE at three representative locations; spread tracks RMSE for approximately 15 time units, indicating that the ensemble faithfully represents its own forecast uncertainty.

\subsubsection{Consistency with ensemble size and EOF truncation}\label{subsec: UQ}

\begin{figure}
\centering
    \begin{subfigure}{1\textwidth}\centering
      \includegraphics[width=1\textwidth]{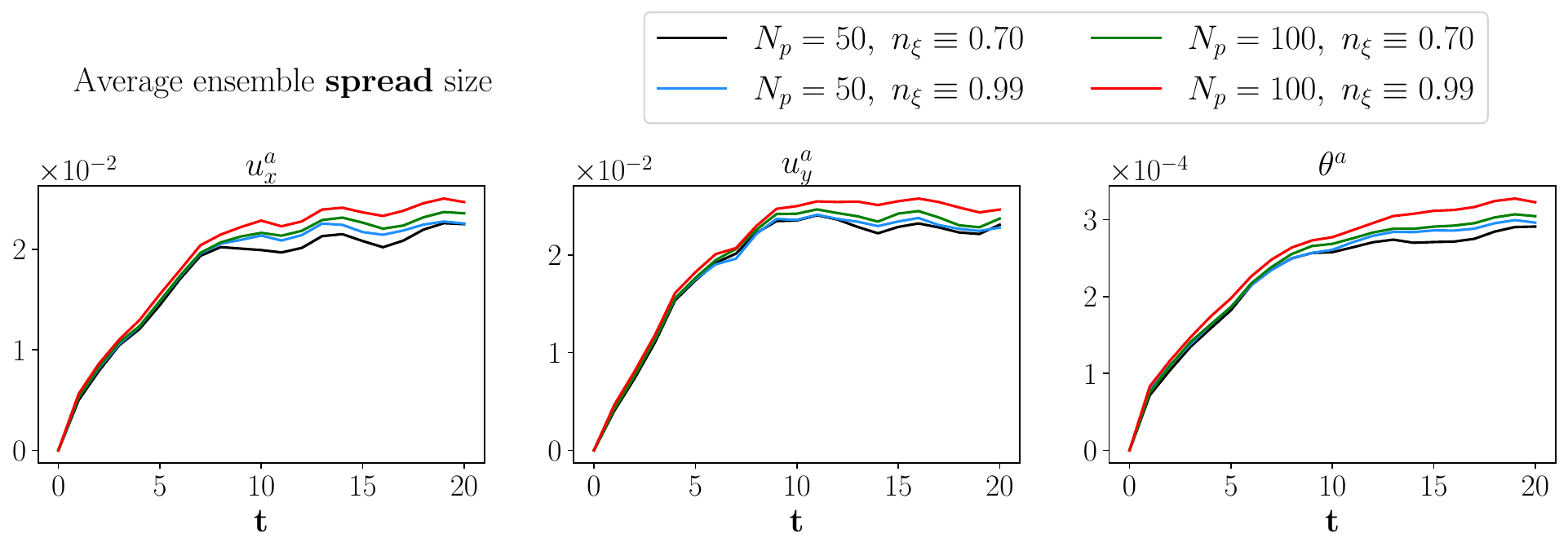}
    \end{subfigure}\par\medskip
    \begin{subfigure}{1\textwidth}\centering
      \includegraphics[width=1\textwidth]{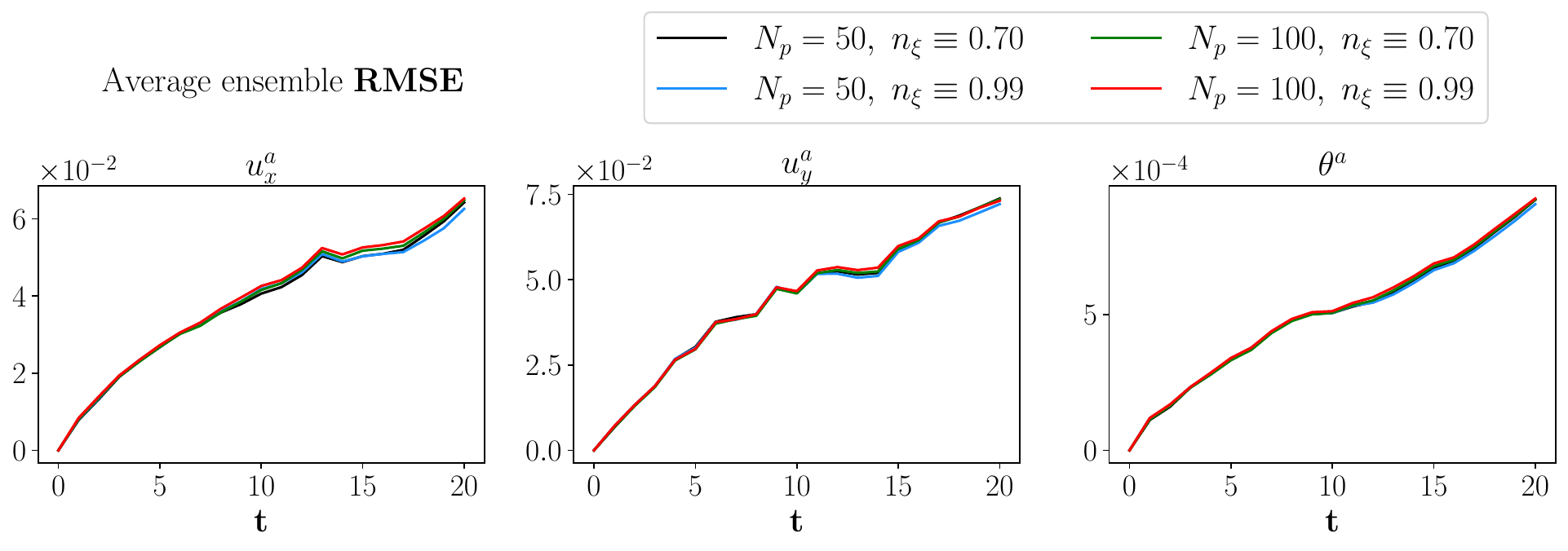}
    \end{subfigure}
    \caption{Domain-averaged ensemble spread (top) and RMSE (bottom) for all four configurations. Spread scales consistently with $N_p$ and $n_\xi$.}
    \label{fig: uq avg rmse spread cm}
\end{figure}

Table~\ref{tab: stoc diff config} lists four configurations varying $N_p \in \{50, 100\}$ and variance threshold $n_\xi \in \{0.70, 0.99\}$. Figure~\ref{fig: uq avg rmse spread cm} shows domain-averaged spread and RMSE for all four. Spread increases monotonically with $N_p$ and $n_\xi$, confirming methodological consistency. The RMSE is less sensitive to these parameters; best RMSE is achieved with $N_p = 50$, $n_\xi = 0.99$, but differences are small. Domain-average spread tracks RMSE for 10--15 time units before RMSE continues growing while spread saturates—a characteristic feature of SALT ensembles that reflects the limited predictability horizon of the coarse model.

\begin{table}[h]
    \centering
    \caption{Stochastic model configurations for UQ experiments.}
    \resizebox{.55\textwidth}{!}{%
    \begin{tabular}{c c c}
    \toprule
        \textbf{Config.} & \textbf{Ensemble size $N_p$} & \textbf{Variance level (\%)}\\
        \midrule
         1 & 50  & 70 \\
         2 & 50  & 99 \\
         3 & 100 & 70 \\
         4 & 100 & 99 \\
         \bottomrule
    \end{tabular}}
    \label{tab: stoc diff config}
\end{table}

\subsubsection{Stochastic versus deterministic ensemble}\label{subsec: stoch vs det ensem}

To compare stochastic and deterministic ensemble strategies directly, both models are run from an ensemble of perturbed initial conditions
$$\mathbf{u}_\text{pert} = \mathbf{u}_0 + 0.2\,r\,\mathbf{u}_0, \quad r \sim \mathcal{N}(0,1),$$
with $N_p = 50$ particles each. The stochastic model uses the OU-calibrated SALT parameterization; the deterministic model propagates each member without any stochastic forcing.

\begin{figure}
\centering
    \begin{subfigure}[b]{\textwidth}\centering
        \includegraphics[width=0.31\linewidth]{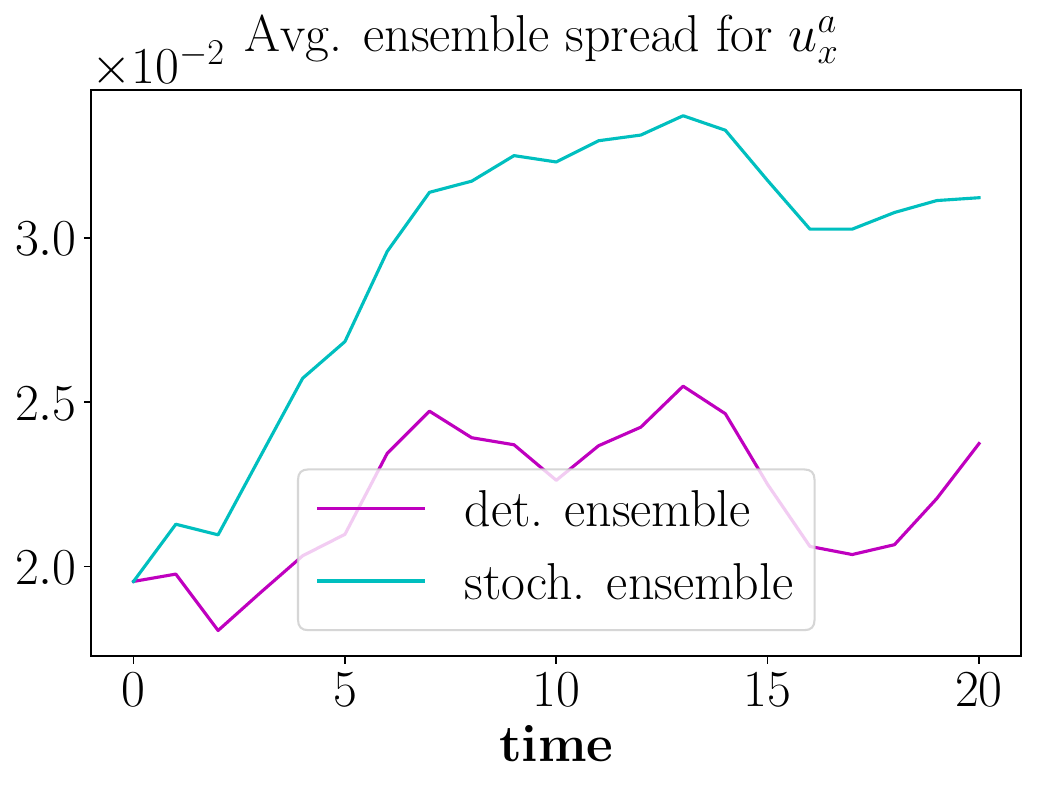}\hfill
        \includegraphics[width=0.31\linewidth]{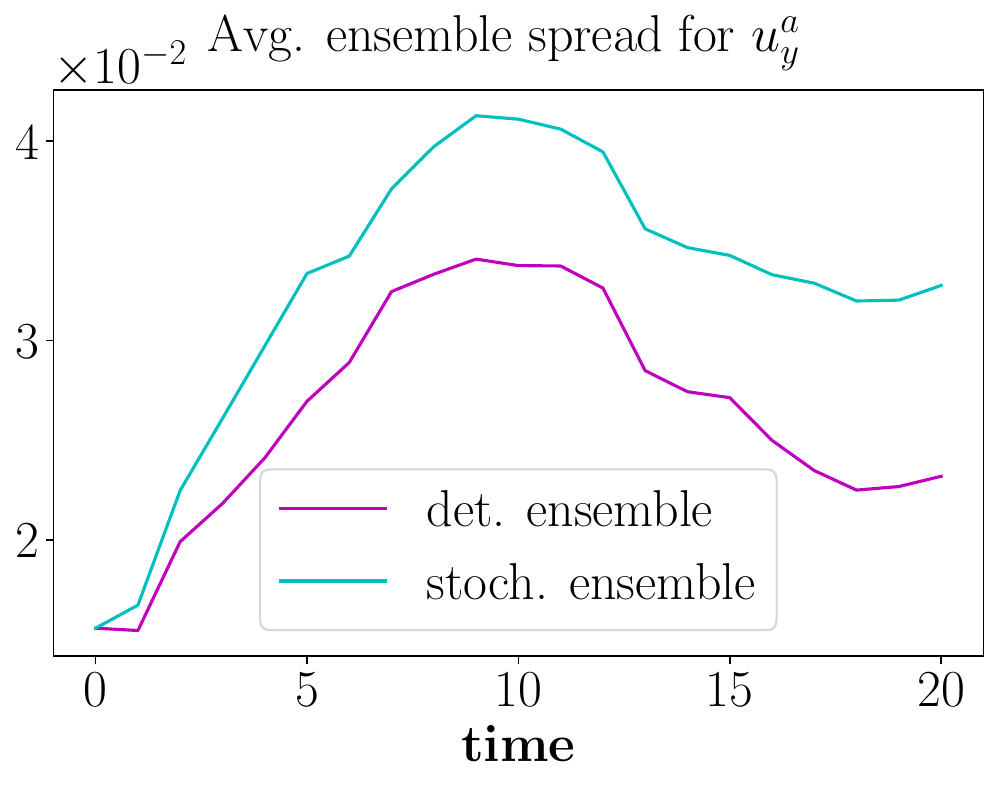}\hfill
        \includegraphics[width=0.31\linewidth]{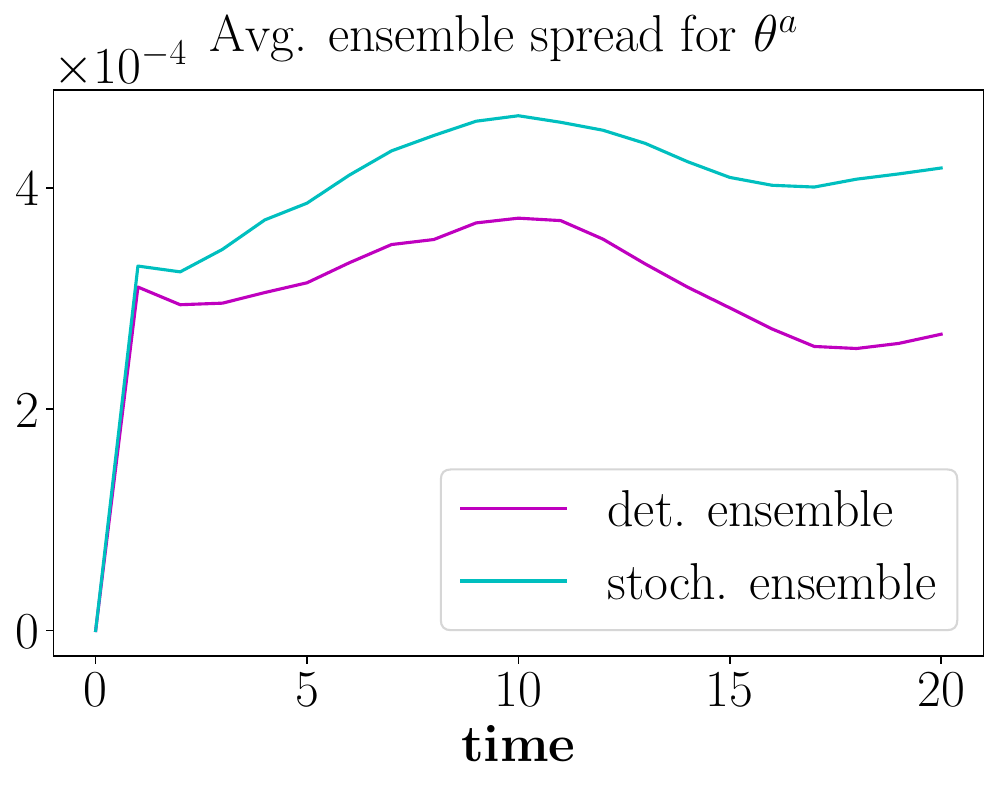}
    \end{subfigure}\par\medskip
    \begin{subfigure}[b]{\textwidth}\centering
        \includegraphics[width=0.31\linewidth]{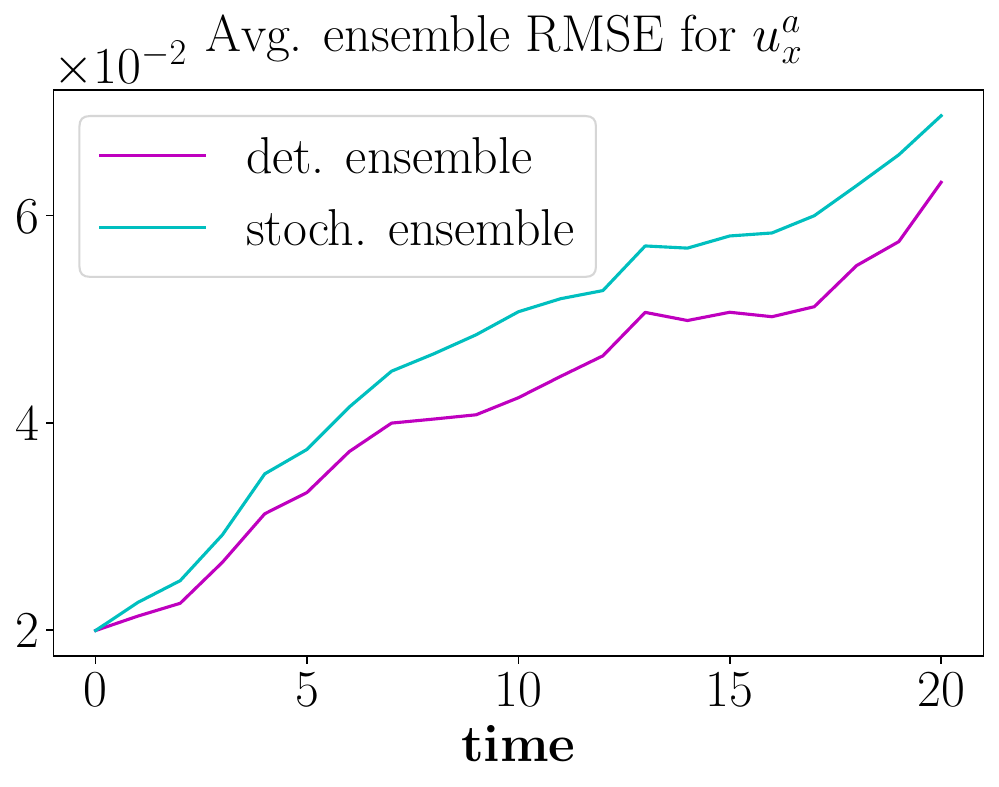}\hfill
        \includegraphics[width=0.31\linewidth]{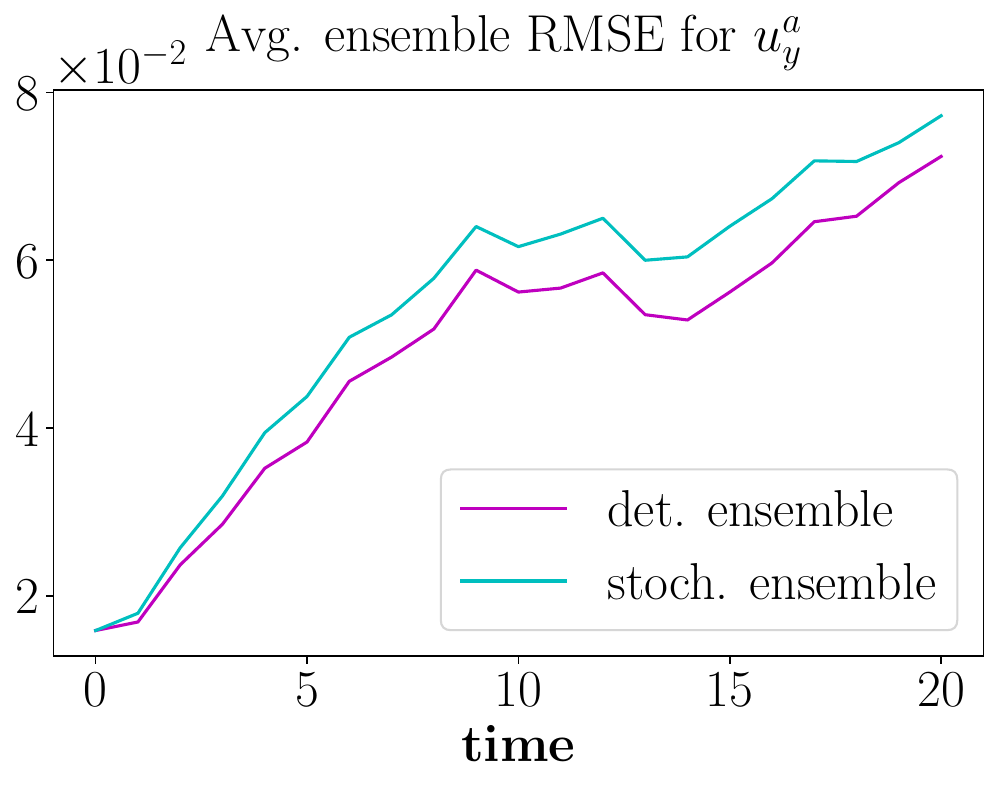}\hfill
        \includegraphics[width=0.31\linewidth]{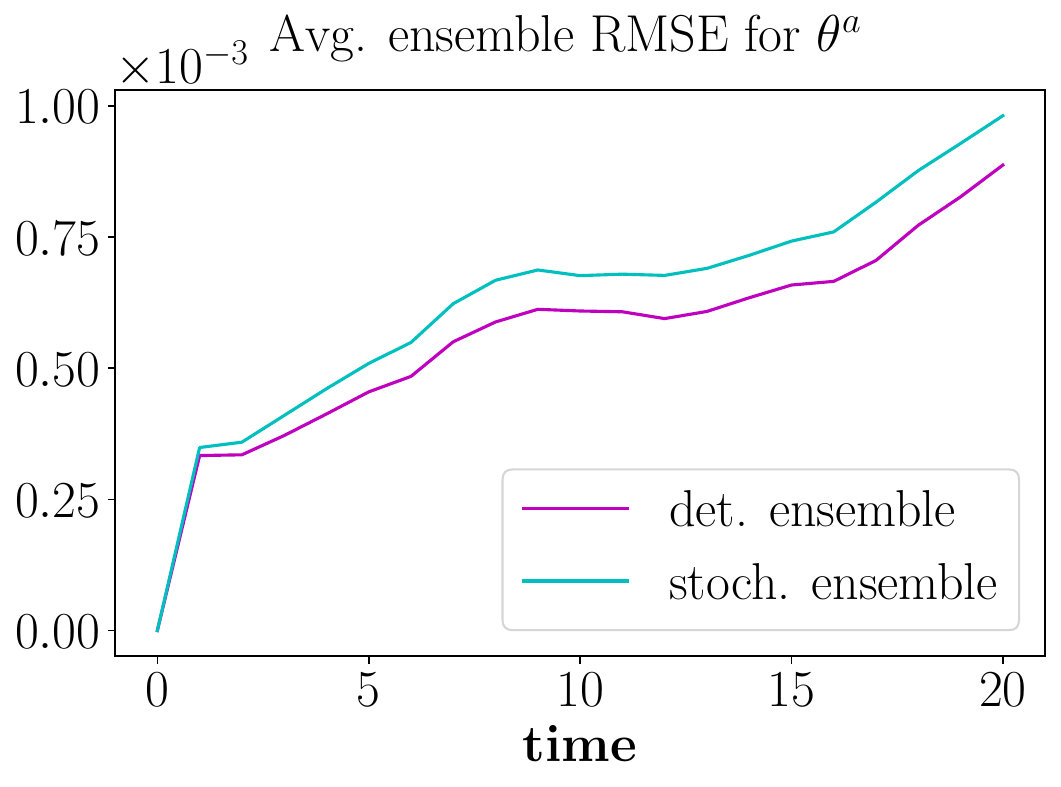}
    \end{subfigure}
    \caption{Domain-averaged spread (top) and RMSE (bottom) for stochastic and deterministic ensembles initialized from perturbed initial conditions. The stochastic ensemble maintains spread--error agreement for 10--12 time units versus 6--8 for the deterministic ensemble.}
    \label{fig: uq avg rmse spread det v stoch cm}
\end{figure}

Figure~\ref{fig: uq avg rmse spread det v stoch cm} compares domain-averaged spread and RMSE for both ensembles. The stochastic model produces consistently larger spread and matches its own RMSE trend for 10--12 time units; the deterministic ensemble spread tracks its RMSE for only 6--8 time units, the well-known under-dispersion of deterministic ensembles \citep{resseguierNewTrendsEnsemble2021}. Paradoxically, the deterministic ensemble has lower RMSE throughout—reflecting higher point accuracy at the cost of insufficient uncertainty representation.

\subsubsection{CRPS and approximation of the forecast distribution}\label{subsec: crps}

A fundamental question for any stochastic parameterization is whether the ensemble correctly approximates the conditional distribution of the true solution given the initial data—that is, whether it samples from the right forecast measure rather than merely tracking the mean. Metrics based on the ensemble mean, such as RMSE or MAE (mean absolute error), cannot answer this question: a perfectly calibrated ensemble with large spread and a tightly clustered but biased ensemble can have identical RMSE while representing entirely different probabilistic forecasts.

The Continuous Ranked Probability Score (CRPS) \citep{hersbachDecompositionContinuousRanked2000, gneitingStrictlyProperScoring2007} addresses this by measuring the discrepancy between the forecast CDF $F$ and the empirical CDF of the verifying observation $y$:
$$\operatorname{crps}(F, y) = \int_{-\infty}^{\infty} (F(x) - \mathds{1}(x \geq y))^2 \,\deriv x.$$
CRPS is a \emph{strictly proper} scoring rule: it is minimized in expectation if and only if $F$ equals the true conditional distribution of $y$ \citep{gneitingStrictlyProperScoring2007}. It therefore rewards both sharpness and calibration simultaneously. In the language of dynamical systems, minimizing the expected CRPS is equivalent to minimizing the Cramér distance between the forecast measure and the true conditional measure of the system—a natural criterion for assessing how well a stochastic model approximates the correct conditional statistics of the underlying flow.

For a finite ensemble of $N_p$ particles sorted as $x_1 \leq \cdots \leq x_{N_p}$, we use the probability-weighted moment estimator \citep{zamoEstimationContinuousRanked2018}:
\begin{equation}\label{eq: crps}
    \widehat{\operatorname{crps}}_\text{PWM}(N_p, y) = \frac{1}{N_p}\sum_{i=1}^{N_p} |x_i - y| + \frac{1}{N_p}\sum_{i=1}^{N_p} x_i - \frac{2}{N_p(N_p-1)}\sum_{i=1}^{N_p}(i-1)x_i.
\end{equation}

\begin{figure}[h]
\centering
    \begin{subfigure}{.48\textwidth}\centering
      \includegraphics[width=1\textwidth]{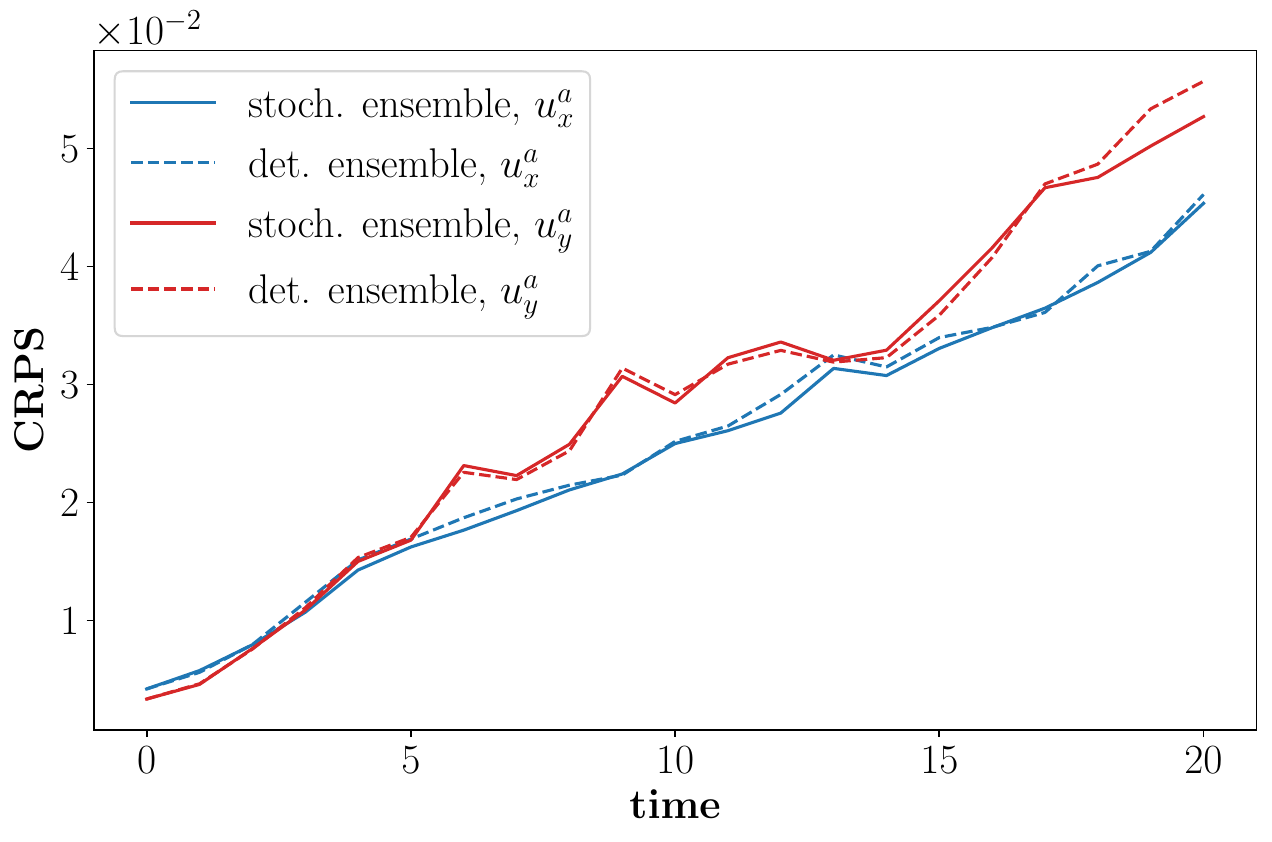}
    \end{subfigure}\hfill
    \begin{subfigure}{.48\textwidth}\centering
      \includegraphics[width=1\textwidth]{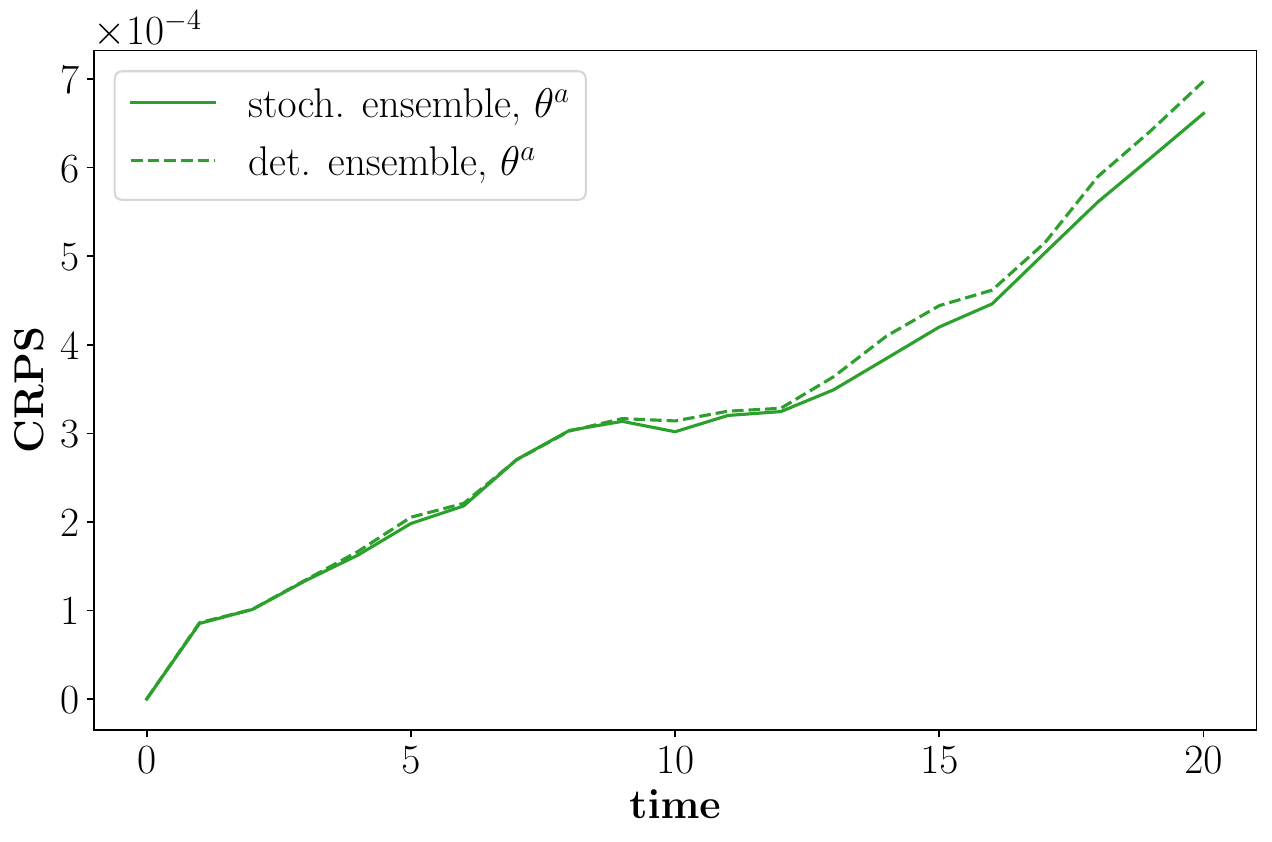}
    \end{subfigure}
    \caption{CRPS for velocity (left) and temperature (right): stochastic ensemble (solid) versus deterministic ensemble (dashed). Despite higher RMSE, the stochastic ensemble achieves lower CRPS throughout, indicating superior probabilistic forecast quality.}
    \label{fig: crps stoch v det ensem}
\end{figure}

Figure~\ref{fig: crps stoch v det ensem} shows CRPS averaged over all 84 observation points at unit intervals from $t=25$ to $t=45$. Despite its lower RMSE, the deterministic ensemble is consistently outperformed by the stochastic ensemble: for $u_y^a$ and $\theta^a$ the stochastic CRPS is lower throughout, and for $u_x^a$ a clear advantage emerges near the end of the simulation. This result has a natural dynamical interpretation: the deterministic ensemble, initialized from perturbed initial conditions but propagated without stochastic forcing, progressively collapses as initial-condition differences are mixed out, ultimately failing to sample the true forecast distribution. The SALT ensemble, by contrast, continues to inject uncertainty through the stochastic transport terms, maintaining a non-degenerate approximation to the forecast measure throughout the simulation window. The CRPS advantage therefore reflects a genuine difference in the ability of the two ensembles to represent the conditional statistics of the coupled system, not merely a calibration artifact.

%%%%%%%%%%%%%%%%%%%%%
\section{Conclusion}\label{sec: conclusion}
%%%%%%%%%%%%%%%%%%%%%

We have presented the first application of the SALT stochastic parameterization framework to a coupled ocean-atmosphere system. The key findings are:

\begin{enumerate}
    \item \textbf{OU noise outperforms white noise.} The leading EOF modes exhibit decorrelation times of 50--150 time steps, making the Gaussian white-noise approximation qualitatively incorrect. Replacing white noise with AR(1)-fitted OU processes—the unique stationary Gaussian Markov model with finite memory—eliminates spurious oscillations and improves ensemble reliability.

    \item \textbf{Ensemble consistency.} The SALT ensemble spread scales correctly with both ensemble size $N_p$ and EOF truncation $n_\xi$, confirming internal consistency of the parameterization and that the methodology correctly samples the subgrid uncertainty.

    \item \textbf{Superior approximation of the forecast measure.} Under CRPS—which measures discrepancy between the forecast measure and the true conditional distribution, and is minimized if and only if the forecast distribution equals the truth—the stochastic ensemble consistently outperforms a size-matched deterministic ensemble initialized from the same perturbed conditions. The spread--error agreement window (10--12 time units) is significantly longer than for the deterministic ensemble (6--8 time units). The CRPS advantage reflects the ability of SALT to maintain a non-degenerate ensemble measure throughout the forecast window, whereas the deterministic ensemble collapses as initial perturbations are mixed out.
\end{enumerate}

Several directions for future work follow naturally. The effect of coarse-grid resolution on SALT performance should be quantified by systematic refinement studies, including a verification of the discrete Kelvin circulation theorem as a diagnostic for structure preservation. Extending the SALT parameterization to the ocean component would fully exploit the coupled framework and allow analysis of how atmospheric stochasticity impacts long-time ocean statistics and the invariant measure of the coupled system. The demonstrated ensemble quality makes this system a natural candidate for combination with particle filter data assimilation \citep{cotterParticleFilterStochastic2020}. Finally, establishing well-posedness of the coupled SALT system \eqref{eq: non dim stoch climate}—existence, uniqueness, and continuous dependence on initial data for the compressible stochastic atmosphere coupled to the incompressible deterministic ocean—remains an important open analytical problem.

\section*{Data availability}
Code, data, and post-processing scripts are available at \url{https://github.com/sharma-kk/coupled_OA_model}. Numerical implementation uses the open-source Firedrake finite element package \citep{rathgeberFiredrakeAutomatingFinite2016}; visualization uses ParaView \citep{ahrensParaViewEndUserTool2005a}.

\section*{Acknowledgments}
K.K.S.\ is grateful to James Woodfield, Sagy Ephrati, and Wei Pan for valuable discussions.

%%%%%%%%%%%%%%%%%%%%
\begin{appendices}
%%%%%%%%%%%%%%%%%%%%

\section{Non-dimensionalization}\label{app: nondim}

The stochastic atmosphere equations in dimensional form read \citep{crisanImplementationHasselmannsParadigm2023}:
\begin{align}\label{eq: stoch atmo}
        \deriv \mathbf{u} &+ \left((\mathbf{u} \deriv t + \sum_i \boldsymbol{\xi}_i \circ \deriv W^i)\cdot \nabla\right) \mathbf{u} + f \hat{\mathbf{z}} \times \left(\mathbf{u}\deriv t + \sum_i \boldsymbol{\xi}_i \circ \deriv W^i_t\right) \nonumber \\
        &+ \sum_i \left(\sum_{j=1}^2 u_j \nabla \xi_{i,j} + \nabla(\boldsymbol{\xi}_i \cdot \mathbf{R})\right)\circ \deriv W^i_t = (-\kappa \nabla \theta + \nu \Delta \mathbf{u}) \deriv t, \\
        \deriv \theta &+ \nabla\cdot\!\left(\theta \!\left(\mathbf{u}\deriv t + \sum_i \boldsymbol{\xi}_i \circ \deriv W^i_t\right)\right) = \eta \Delta \theta \deriv t,
\end{align}
where $\text{curl}(\mathbf{R}) = f\hat{\mathbf{z}}$. Introducing dimensionless variables $\tilde{\mathbf{u}} = \mathbf{u}/U$, $\tilde{\theta} = \theta/\Theta$, $\tilde{x} = x/L$, $\tilde{t} = tU/L$ yields non-dimensional parameters $Ro = U/(Lf)$, $C = U^2/(\kappa\Theta)$. We assume $\boldsymbol{\xi}_i$ divergence-free and neglect the Coriolis noise terms $\hat{\mathbf{z}} \times \sum_i \boldsymbol{\xi}_i \circ \deriv W^i_t + \sum_i \nabla(\boldsymbol{\xi}_i \cdot \mathbf{R})\circ \deriv W^i_t$, replacing physical viscosity with eddy viscosity $\nu_e$ and eddy diffusivity $\eta_e$, arriving at \eqref{eq: non dim stoch climate}.

\section{Estimation of correlation vectors}\label{app: cali}

The correlation vectors $\boldsymbol{\xi}_i$ are estimated from the Eulerian approximation of the Lagrangian trajectory residual:
$$\sum_i \boldsymbol{\xi}_i \circ \Delta B^i_n \approx (\mathbf{u} - \overline{\mathbf{u}})\sqrt{\Delta t}, \quad \Delta B^i_n \sim \mathcal{N}(0,1),$$
where $\overline{\mathbf{u}}$ is obtained by applying the Helmholtz smoothing operator $\overline{\mathbf{u}} - c^2\Delta\overline{\mathbf{u}} = \mathbf{u}$ (with $c$ set to the coarse element size) and projecting onto the coarse grid \citep{cotterParticleFilterStochastic2020, ephratiDataDrivenStochasticLie2023}. The data matrix $F$ of column-mean-centred rescaled residuals is decomposed by SVD: $\operatorname{SVD}(F) = U\Lambda V^T$. The spatial correlation vectors are
$$\boldsymbol{\xi}_i = \frac{\lambda_i}{\sqrt{m-1}}\,\mathbf{v}_i^T,$$
where $\lambda_i$ are singular values and $\mathbf{v}_i$ the right singular vectors, giving orthonormal temporal modes $\sqrt{m-1}\,\mathbf{u}_i$ with unit variance (the EOFs). Full algorithmic details follow \citet{cotterNumericallyModelingStochastic2019, hannachiEmpiricalOrthogonalFunctions2007} and can be found in \citet{sharma2025development}.

\section{AR(1) model for discrete OU processes}\label{app: AR1 model}

Each time series $a_i(t)$ is approximated by an AR(1) process $X_t = \varphi X_{t-1} + \varepsilon_t$, $\varepsilon_t \sim \mathcal{N}(0, \sigma^2)$, which is the discrete analogue of an OU process \citep{mallerOrnsteinUhlenbeckProcesses2009}. The autoregressive coefficient $\varphi$ is estimated as
$$\varphi = \frac{\operatorname{Cov}(a_i(t_k), a_i(t_{k+1}))}{\operatorname{Var}(a_i)} = \frac{1}{M}\sum_{k=0}^{M-1} a_i(t_k)\,a_i(t_{k+1}),$$
and $\sigma = \sqrt{1 - \varphi^2}$ ensures unit stationary variance (Algorithm \ref{alg: AR1}).

\begin{algorithm}
\caption{\label{alg: AR1} Generating discrete OU processes via AR(1)}
        \begin{algorithmic}[1]
        \STATE Given $a_i(t_k)$, $k=0,\ldots,M-1$, with $\operatorname{Var}(a_i)=1$, $\mathbb{E}(a_i)=0$.
        \STATE Compute $\varphi = M^{-1}\sum_{k=0}^{M-1} a_i(t_k)\,a_i(t_{k+1})$.
        \STATE Set $\sigma = \sqrt{1-\varphi^2}$.
        \STATE Initialize $X(t_0) \sim \mathcal{N}(0,1)$.
        \FOR{$j = 0, 1, \ldots, M-1$}
            \STATE $X(t_{j+1}) = \varphi\,X(t_j) + \sigma\,r_j$, \quad $r_j \sim \mathcal{N}(0,1)$.
        \ENDFOR
        \end{algorithmic}
\end{algorithm}

\section{Discretization}\label{app: discretization}

\subsection*{Deterministic model}

The domain is periodic in $x$ with free-slip and insulated Neumann conditions at the $y$-boundaries. The atmosphere velocity and temperature $\mathbf{u}^a, \theta^a \in \mathbb{P}_1$ (piecewise linear) satisfy the variational problem
\begin{align}
    \langle \partial_t \mathbf{u}^a, \mathbf{v}^a\rangle + \langle(\mathbf{u}^a\cdot\nabla)\mathbf{u}^a, \mathbf{v}^a\rangle + \tfrac{1}{Ro^a}\langle \hat{\mathbf{z}}\times\mathbf{u}^a, \mathbf{v}^a\rangle - \tfrac{1}{C^a}\langle\theta^a, \nabla\cdot\mathbf{v}^a\rangle &= -\nu^a\langle\nabla\mathbf{u}^a, \nabla\mathbf{v}^a\rangle, \\
    \langle \partial_t \theta^a, \phi^a\rangle - \langle\theta^a\mathbf{u}^a, \nabla\phi^a\rangle &= \langle\gamma(\theta^a-\theta^o), \phi^a\rangle - \eta^a\langle\nabla\theta^a, \nabla\phi^a\rangle.
\end{align}
Ocean velocity and pressure $(\mathbf{u}^o, p^o) \in \mathbb{P}_2 \times \mathbb{P}_1^\text{disc}$ on a barycentrically refined mesh (Scott--Vogelius elements) satisfy pointwise incompressibility; $\theta^o \in \mathbb{P}_1$. The divergence-free part $\mathbf{u}^a_{sol}$ is obtained via the Helmholtz projection $\mathbf{u}^a_{sol} = \mathbf{u}^a - \nabla q$, where $-\langle\nabla q, \nabla\chi\rangle = \langle\nabla\cdot\mathbf{u}^a, \chi\rangle$, $q \in \mathbb{P}_2$. Crank--Nicolson time integration is used throughout.

\subsection*{Stochastic model}

The stochastic increments $\deriv W^i_t$ are approximated by $w_i^n\sqrt{\Delta t}$, where $w_i^n$ is the AR(1) time series from Algorithm~\ref{alg: AR1}. Defining $\tilde{\mathbf{u}}^n = \sum_i \boldsymbol{\xi}_i w_i^n$, the stochastic atmosphere equations discretize as
\begin{align}
    \langle\mathbf{u}^{n+1}-\mathbf{u}^n, \mathbf{v}\rangle &+ \tfrac{\Delta t}{2}\langle(\mathbf{u}^{n+1}\cdot\nabla)\mathbf{u}^{n+1} + (\mathbf{u}^n\cdot\nabla)\mathbf{u}^n, \mathbf{v}\rangle + \tfrac{\Delta t}{2Ro}\langle\hat{\mathbf{z}}\times(\mathbf{u}^{n+1}+\mathbf{u}^n), \mathbf{v}\rangle \nonumber \\
    &+ \tfrac{\Delta t}{2C}\langle\theta^{n+1}+\theta^n, \nabla\cdot\mathbf{v}\rangle + \tfrac{\nu_e\Delta t}{2}\langle\nabla(\mathbf{u}^{n+1}+\mathbf{u}^n), \nabla\mathbf{v}\rangle \nonumber \\
    &+ \tfrac{\sqrt{\Delta t}}{2}\langle(\tilde{\mathbf{u}}^n\cdot\nabla)(\mathbf{u}^{n+1}+\mathbf{u}^n), \mathbf{v}\rangle + \tfrac{\sqrt{\Delta t}}{2}\langle(u_1^{n+1}+u_1^n)\nabla\tilde{u}_1^n + (u_2^{n+1}+u_2^n)\nabla\tilde{u}_2^n, \mathbf{v}\rangle = 0,
\end{align}
with an analogous equation for $\theta$. The ocean is updated with the ensemble-mean $\overline{\mathbf{u}}^a_{sol}$. Each ensemble member runs on a separate CPU core using Firedrake's parallelization.

\end{appendices}

\bibliography{bibliography.bib}
\end{document}